\documentclass{aa}
\usepackage[varg]{txfonts}
\usepackage{graphicx}
\usepackage{subcaption}
\usepackage[dvipsnames]{xcolor}
\RequirePackage[colorlinks=true,linkcolor=blue,citecolor=blue,urlcolor=blue]{hyperref}
\usepackage{comment}
\usepackage{float}
\usepackage{orcidlink}

\newcommand{\mvir}{{\rm M}_{\rm 500c}}
\newcommand{\rvir}{r_{\rm 500c}}
\newcommand{\mtc}{{\rm M}_{\rm 200c}}
\newcommand{\rtc}{r_{\rm 200c}}
\newcommand{\msun}{M_\odot}
\newcommand{\betaP}{\beta_{\rm P}}
\newcommand{\betaA}{\beta_{\rm A}}

\begin{document}

\title{Magnetic fields in the intracluster medium with TNG-Cluster: properties, morphology, and tangential anisotropy}
\titlerunning{Magnetic Field Morphology in TNG-Cluster}

\author{Katrin Lehle\inst{1}\thanks{E-mail: k.lehle@stud.uni-heidelberg.de} \orcidlink{0009-0005-5700-6483}
\and Mateusz Ruszkowski \inst{2} \orcidlink{0009-0002-2669-9908}
\and Dylan Nelson \inst{1, 3} \orcidlink{0000-0001-8421-5890}
\and Marine Prunier \inst{4, 5, 6} \orcidlink{0009-0003-0932-2487}
\and Annalisa Pillepich \inst{4} \orcidlink{0000-0003-1065-9274}
} 

\institute{Universit\"{a}t Heidelberg, Zentrum f\"{u}r Astronomie, ITA, Albert-Ueberle-Str. 2, 69120 Heidelberg, Germany \label{1}
\and Department of Astronomy, University of Michigan, Ann Arbor, MI 48109, USA \label{2}
\and Universität Heidelberg, Interdisziplinäres Zentrum für Wissenschaftliches Rechnen, INF 205, 69120 Heidelberg, Germany \label{3}
\and Max-Planck-Institut f\"{u}r Astronomie, K\"{o}nigstuhl 17, 69117 Heidelberg, Germany \label{4}
\and Département de Physique, Université de Montréal, Succ. Centre-Ville, Montréal, Québec, H3C 3J7, Canada \label{5}
\and Centre de recherche en astrophysique du Quebec (CRAQ)\label{6}
}

\date{}

\abstract{We characterize the magnetic field properties of 352 massive galaxy clusters from the TNG-Cluster magnetohydrodynamical cosmological simulation with a focus on central magnetic field morphology in cool-core (CC) vs non-cool-core (NCC) clusters. We present the central values and radial profiles of magnetic field strength and plasma parameter as a function of mass, cooling status and redshift. Compared to low-redshift observations, TNG-Cluster produces reasonable magnetic field amplitudes in the central regions of clusters spanning a range of $1-200\, \mu$G. We then discuss the main finding of this work: $z=0$ cool-core clusters have preferentially tangential magnetic fields at a characteristic scale of $\sim 0.1 \rvir$. These strongly tangential field orientations are specific to CCs. In contrast, across the full cluster population, magnetic fields show isotropic configurations at all radii and redshifts. As individual halos grow, the evolution of their magnetic field topologies is diverse: tangential features can be short-lived, persist over large cosmological time-scales, or periodically appear, vanish, and reappear towards $z=0$. We discuss the underlying physics and possible physical scenarios to explain the origin of these structures. We argue that both AGN feedback-driven outflows, and merger-driven sloshing motions, cannot explain the population-wide tangential bias in magnetic field orientation. Instead, we propose that the trapping of internal gravity waves is responsible for the tangentially biased magnetic field topologies that we find in cool-core TNG-Cluster halos, due to the strong entropy gradient in these clusters.}


\maketitle


\section{Introduction}
\label{sec_intro}

Magnetic fields pervade the Universe and play an important role in many astrophysical phenomena. Galaxy clusters provide a particularly rich environment for magnetic field physics. Observations infer that the intracluster medium (ICM) is magnetized, with volume-filling magnetic fields permeating this plasma. The ICM is visible through extended and diffuse synchrotron emission in the form of radio halos. Magnetic fields in clusters are also probed by Faraday rotation measure sightlines towards background objects \citep{kim1991, giovannini1999, carilli2002, govoni2012}. The magnetic field strength in the ICM is typically $\sim \mu$G in strength, with order of magnitude scatter between different measurement methods, across individual clusters, and within clusters \citep{carilli2002, govoni2012}.

Although magnetic fields are not dynamically dominant in the volume-filling ICM, they can play a significant role in shaping its microscopic physics. Magnetic fields suppress transport processes perpendicular to field lines, affecting heat conduction \citep{kannan2016, talbot2024}, the dynamics of cosmic rays \citep[CRs;][]{wiener2019,vazza2021} as well as their observable signatures \citep{pfrommer2008}. They can also suppress Kelvin-Helmholtz instabilities and can promote thermal instabilities, which amplify overdensities in the gas \citep{ruszkowski2023}. In addition, magnetic tension impacts the buoyancy of cooling gas \citep{ehlert2018}, implying that cooling flows and condensing cold clouds are more resilient against disruption \citep{ramesh2024}, while the conditions and onset of thermal instability also change \citep{wibking2025}. Although sub-dominant, magnetic pressure adds an additional, non-thermal pressure component to the ICM \citep{goncalves1999,dolag2000}. Finally, non-ideal MHD effects including finite resistivity \citep{bonafede2011}, and ohmic and ambipolar diffusion \citep{zier2024} can have additional impact, while interactions with cosmic rays can influence magnetic field amplification \citep{marcowith2016}.

Despite the detection and measurement of ICM magnetic field strengths in clusters, their detailed morphology and amplification processes remain unclear. The resulting magnetic field morphology can be complex. Spatially resolved rotation measure (RM) mapping, although sparse, suggests that magnetic fields tend to be tangled rather than regularly ordered, with coherence scales of only $5-30\,$kpc \citep{carilli1991, carilli2002}. At the same time, recent radio observations show the existence of thin synchrotron-emitting filaments in the ICM \citep{rudnick2022,brienza2025}. Potential causes include both collimated jets from AGN as well as stripped tails of jellyfish galaxies, reflecting the complex interactions and resulting structure of magnetic fields in clusters.

The relationship between magnetic field properties and the thermodynamic state of their host galaxy clusters is also poorly understood. In particular, the central states of galaxy clusters are heterogeneous and form a continuum from cool-core (CC) clusters to non-cool-core (NCC) clusters \citep{molendi2001,lehle2024}. CC clusters have low central cooling time and entropy and are characterized by a prominent surface brightness peak because of high central density. NCCs are the opposite in terms of all these features, and in general the origin of CC versus NCC clusters and the evolutionary pathways between these states remain unclear \citep{allen2001,mccarthy2004,lehle2025}. 

With respect to magnetic fields, CCs and NCCs also differ. Magnetic field strengths in CCs are larger by factors of $\sim 2-3$ than in NCCs \citep{carilli2002}. In addition, coherence lengths are larger for NCCs ($15-30\,$kpc) than for CCs \citep[$5-10\,$kpc;][]{eilek2002}. Using the gradient technique, \citet{hu2020} probe the plane-of-the-sky orientation of magnetic fields in four galaxy clusters, finding that they follow sloshing arms in the Perseus cluster and in M87. In the three CCs of the sample, they also find that the mean orientation of magnetic fields is preferentially tangential. On the other hand, the magnetic fields in the central region of the Coma (NCC) cluster do not appear to have a preferred direction.

From the theoretical perspective, cluster magnetic fields can be studied with idealized halo simulations and in cosmological simulations (using large volumes or zoom-ins). In all cases, magnetic fields must originate from a non-vanishing seed field at early times. Several mechanisms have been proposed to explain the origin of seed magnetic fields \citep{subramanian2016, donnert2018}. Due to the uncertainty in the (primordial) seeding mechanism at high redshift, there are only weak constraints on the seed field strength, of roughly $\sim10^{-34} - 10^{-10}\,$G \citep{donnert2018}. Subsequently, magnetic fields in the ICM can be amplified in-situ as a result of several physical processes.

Due to the conservation of magnetic flux in ideal magnetohydrodynamics (MHD), magnetic field strength increases due to adiabatic compression, with amplitude $B\propto\rho^{2/3}$ that increases with gas density (assuming isotropic turbulence). In the central ICM of clusters, gravitational-sourced contraction can lead to magnetic field strengths $\sim 0.1\,\mu\mathrm{G}$ \citep{dubois2008}. In addition, amplification via turbulence due to the small-scale dynamo is rapid and effective \citep{donnert2018, roh2019, steinwandel2022}. In a turbulent, ionized plasma, the kinetic and magnetic energies are coupled via back-reaction of the field on the flow. As a result, turbulent kinetic energy is converted into magnetic energy in a process that can be thought of as the stretching and folding of existing field lines. Random motions in turbulent flow repeatedly stretch the field lines, and flux conservation ensures that this process locally amplifies the magnetic field strength \citep{biermann1951, vajnshtejn1972, schekochihin2001}.

Non-radiative magneto-hydrodynamical cosmological simulations indicate that the turbulent dynamo can amplify seed fields to \(\mu\mathrm{G}\) in clusters \citep{vazza2014, vazza2017}. In general, the turbulent dynamo successfully operates in cosmological simulations \citep{pakmor2017, pakmor2024, steinwandel2022}, although the small-scale dependence may make the amplification sensitive to numerical resolution \citep{rieder2016,martin-alvarez2018}. In addition to in-situ amplification, magnetic fields can be brought in as part of gravitational collapse and structure formation. At lower redshifts, stellar and AGN feedback can also transport magnetic fields from galactic to halo scales \citep{dubois2008, ramesh2023}. In particular, AGN-driven outflows can be a significant source of ICM-scale magnetic fields \citep{xu2009,aramburo-garcia2021}. Notably, magnetic field strengths are stronger than expected from simple adiabatic compression when baryonic processes such as cooling and feedback, and the resulting galactic scale outflows, are included \citep[e.g.][]{marinacci2015,marinacci2018}. 

In this work, we explore the properties of magnetic fields in the TNG-Cluster simulation \citep{nelson2024}. We focus on magnetic field morphology and the differences between cool-core (CC) and non-cool-core (NCC) clusters. TNG-Cluster is a cosmological magnetohydrodynamical simulation that samples several hundred high-mass galaxy clusters with self-consistently modeled magnetic fields. Notably, TNG-Cluster has been shown to have broadly realistic ICM properties, from total gas content, X-ray luminosities and SZ y-parameters \citep{nelson2024}, to CC statistics \citep{lehle2024}, satellite galaxy populations \citep{rohr2024a}, radio relic emission \citep{lee2024}, AGN-inflated and X-ray visible bubbles and cavities \citep{prunier2025, prunier2025a}, multi-phase kinematics \citep{ayromlou2024}, and core turbulence consistent with recent XRISM observations \citep{truong2024,xrismcollaboration2025a}.

In this work, we have three goals. First, we give a broad census of magnetic field properties in the TNG-Cluster simulation. We then identify a striking difference in the morphology of magnetic fields in CC vs NCC clusters. Finally, we present a plausible physical explanation for this difference. The remainder of this paper is structured as follows. Section~\ref{sec_TNG-Cluster} introduces the TNG-Cluster, Section~\ref{sec_simsBfield} describes details of the magnetohydrodynamics, and Section~\ref{sec_defs} summarizes our methodological choices. In Section~\ref{sec_Bfields} we present the general magnetic field properties of the TNG-Cluster population, while Section~\ref{sec_morphology} examines the morphology of magnetic fields across the whole sample. In Section~\ref{sec_discussion} we discuss the implications of our work and review possible explanations for our findings in Section~\ref{sec_results}. Section~\ref{sec_conclusions} concludes with a summary of our findings.

\section{Methods} \label{sec_methods}

\subsection{The TNG-Cluster simulation} \label{sec_TNG-Cluster}

The TNG-Cluster project consists of 352 high-resolution zoom simulations of the most massive and rare galaxy clusters.\footnote{\url{www.tng-project.org/cluster}} It is an extension of the IllustrisTNG project \citep[hereafter TNG;][]{nelson2018, pillepich2018a, marinacci2018, springel2018, naiman2018, pillepich2019, nelson2019b}, a suite of cosmological gravo-magnetohydrodynamical simulations of galaxy formation and evolution chiefly consisting of the flagship runs TNG50, TNG100, and TNG300.

These simulations are run with the AREPO code \citep{springel2010}, which solves the coupled equations for self-gravity and ideal magnetohydrodynamics \citep{pakmor2011, pakmor2013}. A key feature of the TNG simulations is its robust and well-validated physical model for galaxy formation and evolution, as described in \cite{weinberger2017} and \cite{pillepich2018}. TNG-Cluster utilizes the same unchanged model, which incorporates the key processes relevant to the formation and evolution of galaxies and galaxy clusters, including heating and cooling of gas, star formation, evolution of stellar populations and chemical enrichment, stellar feedback, as well as growth, merging and multi-mode feedback from SMBHs. TNG-Cluster adopts the fiducial TNG cosmology, consistent with \cite{planckcollaboration2016}: $\Omega_m = 0.3089$, $\Omega_b = 0.0486$, $\Omega_\Lambda = 0.6911$, $H_0 = 100 h$\,km s$^{-1}$Mpc$^{-1}$ = 67.74\,km\,s$^{-1}$Mpc$^{-1}$, $\sigma_8 = 0.8159$, and $n_s = 0.9667$.

TNG-Cluster is an extension of the TNG300 simulation by providing better sampling and statistics for highest-mass halos. The target clusters for the zoom (re-)simulations were selected from a large (1\,Gpc)$^3$ dark matter-only run. Halos were chosen exclusively based on their mass at $z=0$: all halos with $\log_{10} (\mtc/\msun) > 15.0$\footnote{$\mtc$ is the mass enclosed within $\rtc$, which is the radius enclosing a sphere with average density 200 times higher than the critical density of the universe at a given redshift. $\rvir$ is the radius enclosing a sphere with average density 500 times higher than the critical density of the universe at a given redshift, and $\mvir$ is the mass enclosed in $\rvir$.} were included, and for $14.5 < \log_{10} (\mtc/\msun) < 15.0$, halos were randomly selected in 0.1 dex mass bins to give a uniform distribution and account for the drop-off in high-mass halos in TNG300 \citep[see Figure 1 of][] {nelson2023}. The resolution of TNG-Cluster matches that of TNG300-1, with gas cell masses of $1.2\times10^7 \msun$ and dark matter particle masses of $6.1 \times 10^7 \msun$.

Halos are identified using the standard friends-of-friends (FoF) algorithm with a linking length of $b=0.2$. Substructures are identified with the \textsc{SUBFIND} routine \citep{springel2001} and their connections across snapshots are tracked using the \textsc{SubLink} algorithm \citep{rodriguez-gomez2015}.

\subsection{Magnetic fields in TNG-Cluster}\label{sec_simsBfield}

In TNG-Cluster, magnetic fields are self-consistently simulated by solving the coupled system of equations of ideal magnetohydrodynamics \citep{pakmor2011}:
\begin{equation}
  \label{eq:cons_eqn}
  \frac{\partial \vec{U}}{\partial t} + \vec{\nabla} \cdot \vec{F} = 0
\end{equation}

\noindent
where $\vec{U}$ is the state vector of conserved quantities and $\vec{F}$ is the vector of corresponding fluxes:
\renewcommand*{\arraystretch}{1.5}
\begin{equation}
  \label{eq:conv_eqn}
  \vec{U} = \left( \begin{array}{c} \rho \\ \rho \vec{v} \\ \rho e \\ \vec{\rm{B}} \end{array} \right),
  \,\,
  \vec{F} = \left( \begin{array}{c} 
      \rho \vec{v} \\
      \rho\vec{v}\vec{v}^T + \rm{P}_{\rm{tot}} - \vec{\rm{B}}\vec{\rm{B}}^T \\
      (\rho e + \rm{P}_{\rm{tot}}) \vec{v} - \vec{\rm{B}} \left( \vec{v} \cdot \vec{\rm{B}} \right) \\
      \vec{\rm{B}}\vec{v}^T - \vec{v}\vec{\rm{B}}^T
    \end{array} \right)
\end{equation}
\noindent
Here, $\rho$, $\vec{v}$, and $\vec{B}$ are the gas density, velocity, and  magnetic field, respectively. Furthermore, the total gas pressure is given by $\rm{P}_{\rm{tot}} = \rm{P}_{\rm{gas}} + \frac{1}{2}\vec{B}^2$ and the total energy per unit mass by $e = u + \frac{1}{2} \vec{v}^2+ \frac{1}{2\rho}\vec{B}^2$, with $u$ specifying the thermal energy per unit mass. 
These equations reduce to the equations of ideal hydrodynamics for the case $\vec{B} = 0$ \citep[as in][]{springel2010}.

A detailed explanation of the numerical techniques and implementation of MHD in AREPO can be found in \cite{pakmor2011, pakmor2013}. Here, we only give an overview of the most important aspects. 

To satisfy the divergence-free constraint, a Powell 8-wave cleaning scheme is adopted in order to control divergence errors at a reasonable level \citep{powell1999,pakmor2013}. The approach incorporates additional source terms into the momentum, induction, and energy equations, enabling the passive advection of $\nabla \cdot \mathbf{B}/\rho$ along the flow. These terms effectively suppress the growth of local $\nabla \cdot \mathbf{B}$ errors. Based on practical experience, this scheme is highly stable and robust. An additional advantage is its locality, as it does not impose extra constraints on the time-step, even when individual time-stepping is employed for all cells \citep{pakmor2013}.

In TNG-Cluster magnetic fields are seeded homogeneously with an initial value of $10^{-14}$ comoving Gauss. This is the same choice as for all IllustrisTNG simulations. It has been shown in previous work that the final outcome within collapsed halos is insensitive to the initial value chosen across several orders of magnitude \citep{ marinacci2015,pakmor2017,aramburo-garcia2021}.

\begin{figure*}
    \centering
    \includegraphics[width=0.9\textwidth]{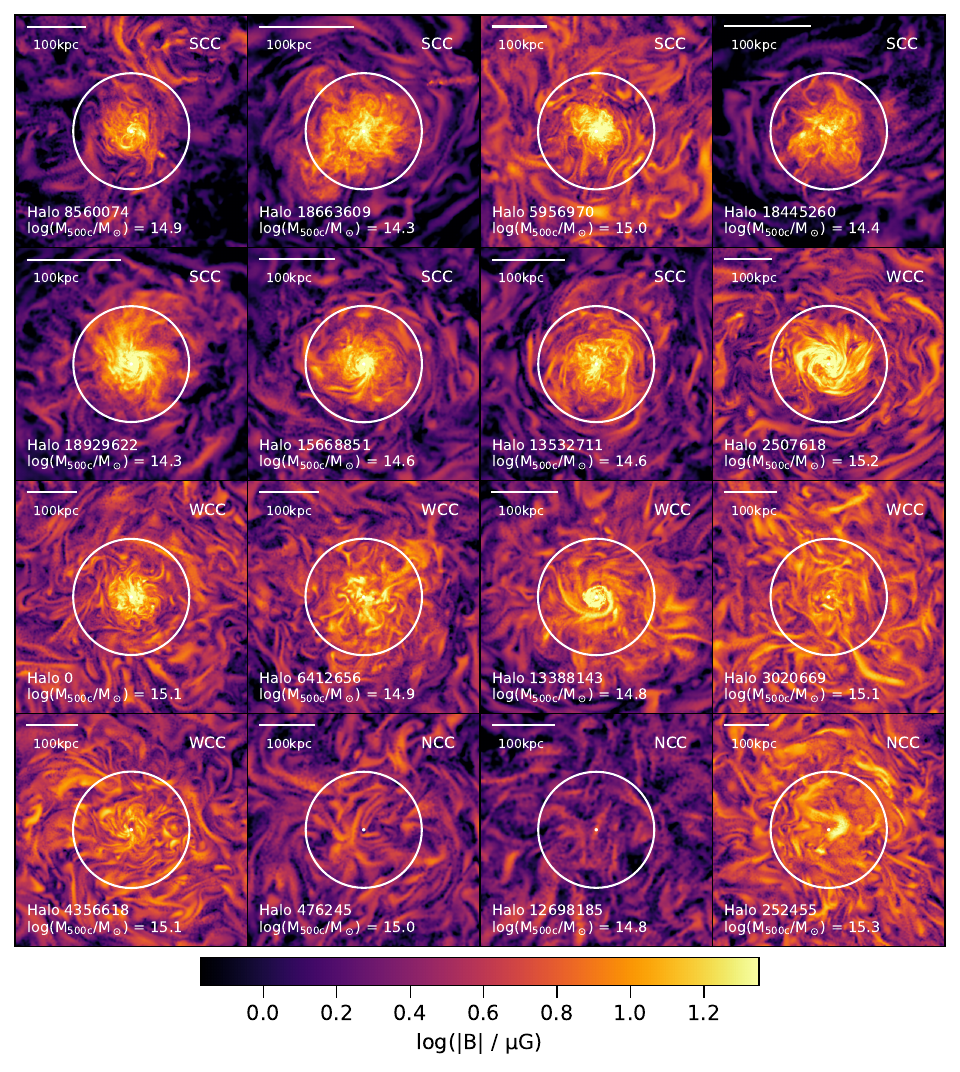}
    \caption{Gallery of magnetic field strength and morphology in sixteen high-mass galaxy clusters selected from TNG-Cluster at $z=0$. Each panel shows the central region and extends $0.2\rvir$ from side-to-side, giving projections of mean magnetic field strength in a thin slice of $15$\,kpc depth. The white circles indicate $0.1\rvir$. The panels are ordered by ascending central entropy $K_0$, such that the strongest cool-core (CC) systems are in the upper left, and the strongest non-cool-core (NCC) halos are located in the bottom right. Central magnetic field strength increases rapidly with halo mass, and is higher in CC versus NCC clusters.}
    \label{fig:galleryB}
\end{figure*}

In the ideal MHD cosmological simulations of structure formation with the AREPO code, magnetic fields undergo efficient amplification due to a combination of gravitational collapse and galactic feedback processes. In regions of low density, such as voids and filaments, magnetic fields grow according to flux conservation. In contrast, in and around halos, where densities are much higher and astrophysical phenomena are more complex, additional effects like turbulence and shear flows, induced by cosmic structure formation, stellar and AGN feedback, become important. As a result, the magnetic field experiences rapid exponential growth driven by small-scale dynamo processes within halos. These processes amplify magnetic fields far beyond what flux conservation alone would predict, reaching strengths of about $\sim 10\,\mu\rm G$ in the centers of cluster-mass halos \citep[][using TNG300]{marinacci2018}. At this mass scale and numerical resolution, most of the magnetic field amplification is already complete by $z \simeq 2$.

In particular, within halos, the shape of the magnetic power spectra and its evolution are consistent with the expectations of a turbulent dynamo \citep{pakmor2024}. This initial phase is followed by a slower, linear amplification stage \citep{pakmor2017, pakmor2020}. Within galaxies themselves, the magnetic field energy tends to saturate at roughly equipartition with the thermal energy. On the other hand, in the circumgalactic medium of galaxies, magnetic field pressure is subdominant in the volume-filling hot phase \citep{nelson2020}.

\subsection{Quantifying clusters and their magnetic fields} \label{sec_defs}

In this study, we examine the 352 galaxy clusters that form the primary sample of TNG-Cluster. At $z=0$ these clusters span $\mvir = 1 \times 10^{14}\,\msun$ up to $\mvir = 1.9 \times 10^{15}\,\msun$, with an average mass of $\mvir = 4.3 \times 10^{14}\,\msun$. We characterize their properties and magnetic field morphologies as follows.

To quantify the morphology of the magnetic fields we compute the magnetic anisotropy $\betaA$:
\begin{equation} \label{eq:betaA}
     \beta_{\rm A} = 1 - \frac{\langle B_{\rm t}^2\rangle_{\rm V}}{2\langle B_{\rm r}^2\rangle_{\rm V}} \, .
\end{equation}
where $B_{\rm r}$ and $B_{\rm t}$ are the radial and tangential components of the local magnetic field, and we adopt a volume weighting. Directly analogous to the $\beta$-anisotropy parameter for stellar orbits \citep{binney1987}, $\betaA$ quantifies the relative strength of the radial to tangential components of the magnetic field. If $\betaA >0$, the magnetic field lines are predominately radially oriented; if $\betaA\sim0$, they are mainly isotropic; and if $\betaA<0$, the magnetic field lines are preferably tangentially oriented (i.e., they are perpendicular to the radial direction). 

We calculate the thermal pressure as $P_{\rm th} = 2n_e k_B T$ and the magnetic pressure as $P_{\rm mag} = \frac{B^2}{8\pi}$. The plasma beta parameter then quantifies the ratio of thermal to magnetic pressure in the gas. To avoid ambiguity, we label the plasma beta parameter by $\beta_{\rm P} = P_{\rm th}/ P_{\rm mag}$ and the anisotropy parameter as $\betaA$, as both parameter are commonly denoted by $\beta$ in the literature. 

We characterize the cooling state of clusters by their central entropy, computed as $K = k_{\mathrm{B}} T n_{\mathrm{e}}^{-2/3}$. The central entropy, $K_0$, is determined using all gas within a three-dimensional aperture of $r<10$\,kpc, centered on the gravitational potential minimum. The calculation includes only gas that satisfies the following criteria: (i) non-star-forming, (ii) net cooling, and (iii) temperature $T>10^6$\,K. These conditions approximately select the X-ray emitting gas typically used in observational estimates of $K_0$. We adopt the observationally motivated thresholds of \cite{hudson2010}. Strong cool-core clusters (SCCs) have $K(r<10$\,kpc$) \equiv K_0 \leq 22$\,keV\,cm$^2$, weak cool-cores (WCCs) are clusters with 22\,keV\,cm$^2 < K_0 \leq 150$\,keV\,cm$^2$, and non-cool-cores (NCCs) have $K_0> 150$\,keV\,cm$^2$ \citep[see][for further discussion on thresholds and classification]{lehle2024}.


\section{Results on magnetic fields from TNG-Cluster} \label{sec_results}
\subsection{Magnetic field strength at z=0 and global morphology in the intracluster medium} \label{sec_Bfields}

\begin{figure*}
    \centering  \hspace*{-1cm}\includegraphics[width=0.75\textwidth]{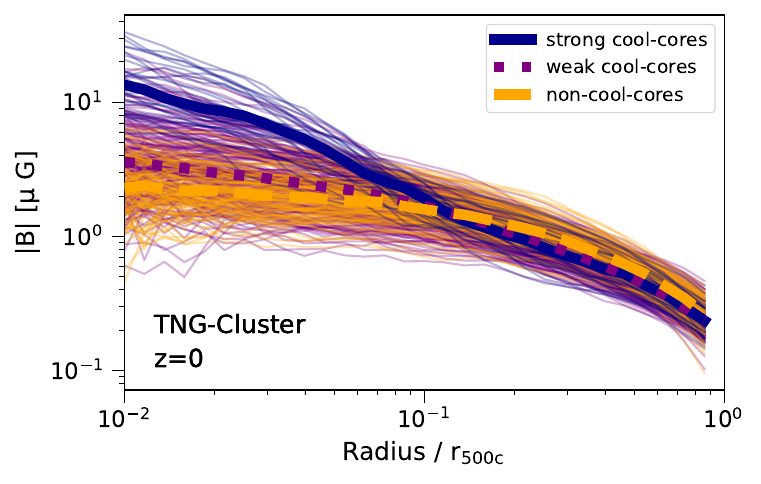}
    \includegraphics[width=0.49\textwidth]{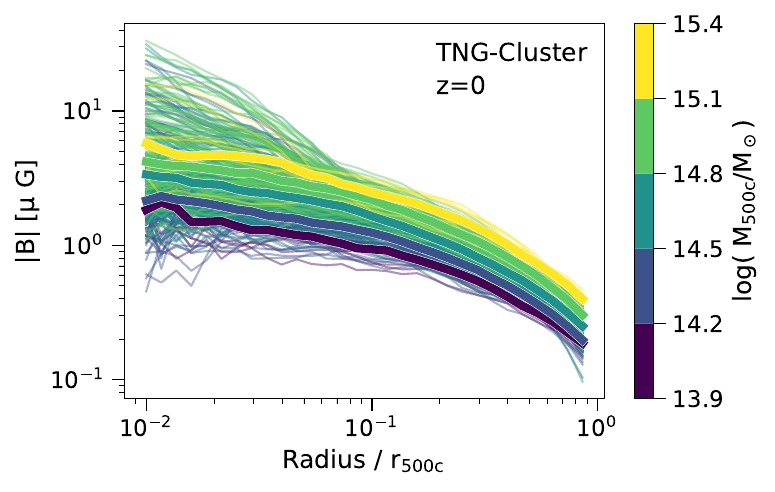}
    \includegraphics[width=0.49\textwidth]{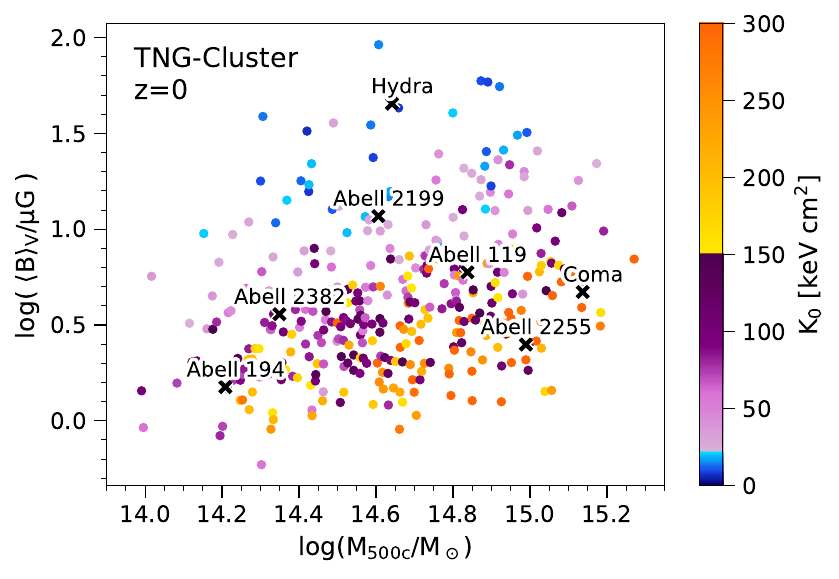}
    \caption{\textbf{Main panel:} radial profiles of the magnetic field strength at $z=0$ for all TNG-Cluster halos, colored according to central entropy, i.e., cool-core versus non-cool-core state. Thin lines show individual clusters, while the three thick lines show median stacks of SCC, WCC, and NCC clusters. \textbf{Lower left:} the same radial profiles of magnetic field strength, except coloring and stacking is based on halo mass. \textbf{Lower right:} trend of volume-average magnetic field strength within $0.1 \rvir$ as a function of halo mass, where color shows central entropy. For comparison, we include central magnetic field strengths from observations \citep[see][and references therein]{govoni2017}. SCCs exhibit distinct radial profiles, with a strong increase in magnetic field strength at $ \lesssim 0.1 \rvir$, and at fixed mass, SCCs have higher central magnetic field strengths compared to WCCs and NCCs.}
    \label{fig:profilesB}
\end{figure*}

We begin our population-level analysis of magnetic field properties in TNG-Cluster by visualizing the magnetic field strength at $z=0$ in the center of sixteen halos (Fig.~\ref{fig:galleryB}). The halos were chosen based on thermodynamic properties to illustrate the great diversity of cluster cores \citep{lehle2024}. Each panel depicts the central $0.2\rvir$ of each halo in a thin slice of depth 15\,kpc. The halos are ordered by central entropy, such that the strongest CC systems are in the upper left corner.  Fig.~\ref{fig:galleryB} clearly shows that the magnetic field strength varies on small scales, indicating highly tangled magnetic fields. The magnetic field is of the order of several $\mu \rm G$, reaching values between $1 - 25 \,\mu \rm G$. For most of the clusters, the magnetic field strength increases toward the center, and this increase is stronger for CCs than for NCCs. The maps also reveal that the overall magnetic field strength reaches higher values for more massive clusters. The mass dependence is the strongest at $r>0.1\rvir$ and less visible inside this radius. 

Diverse phenomena throughout cluster histories can influence both the strength and morphology of magnetic fields in cluster cores. For instance, in the spiral-like gas distribution shown in the last panel of the second row, the spiral features exhibit a strong magnetic field extending to large radii. In the upper-left panel (Halo 8560074), a merger is occurring between two halos with a mass ratio of $M_{\rm main}/M_{\rm sub} \simeq 150$ \citep{lehle2024}, where the magnetic field wraps around the infalling halo core. This halo also has X-ray cavities attached to the SMBH \citep{prunier2025}, around which magnetic fields are draped. In the next panel to the right (Halo 18663609), a satellite is falling into the cluster, leaving behind a tail of stripped material and generating a bow shock clearly visible in density maps \citep{lehle2024}. The magnetic field strength map reveals that this tail has a significantly higher magnetic field strength compared to the surrounding ICM \citep[see also][]{werhahn2025,kurinchi-vendhan2025}.

Fig.~\ref{fig:profilesB} shows radial profiles of magnetic field strength for all halos in TNG-Cluster, split by CC status (main panel) or by halo mass (lower left panel). On average, magnetic field strength increases toward cluster cores. For $r>0.1\rvir$, the scatter in magnetic field strength is relatively small among individual clusters, but it increases significantly at smaller radii. In the outskirts at $\sim \rvir$, the magnetic field strength is approximately $0.3\,\mu\rm G$, while in the core, the scatter ranges from $1\,\mu\rm G$ to $30\,\mu\rm G$.

These findings are in agreement with previous work with the TNG300 simulation \citep{marinacci2018}. They find similar field strengths and an increase in magnetic field strength toward the core, arguing that the field in the core gets amplified by the combined effects of radiative cooling (i.e. increased gas densities) and AGN feedback, which induces a lot of turbulent motions and shear in the center of clusters.

With respect to CC status, we see that for $r>0.1\rvir$, the magnetic field strength profiles of CC and NCC clusters are similar. However, at smaller radii, SCC clusters begin to diverge, showing much higher magnetic field strengths compared to their WCC and NCC counterparts. In terms of mass dependence, magnetic field strength is, on average, higher in more massive halos at all radii. The median profiles, binned by $z=0$ halo mass $\mvir$, are well-ordered by mass and have a consistent shape across all bins. Notably, clusters with steep increases in $|B|$ near the core are present across the entire range of cluster masses.  

In order to assess the mass dependency and the dependency of the CC status in the cluster center we show in the lower right plot of Fig.~\ref{fig:profilesB} the volume-weighted average magnetic field strength within $r<0.1\rvir$ as a function of $z=0$ halo mass and central entropy (color). Magnetic field strength increases significantly with halo mass, ranging from $ \sim 1 \, \mu\mathrm{G}$ for the lowest-mass halos to $\sim 10 \, \mu\mathrm{G} $ for the most massive halos. A strong correlation with CC status, represented by central entropy, is also evident. SCC clusters occupy the upper-right corner of the plot: at fixed mass, they exhibit higher magnetic field strengths, and at fixed field strength, they correspond to lower mass halos. The opposite trend is true for NCC clusters.

We make a qualitative comparison to some available observational measurements of central magnetic field strength in clusters \citep{govoni2017}. These are marked with black crosses and labeled with individual cluster names. Hydra, for example, falls squarely in our CC regime, while the most massive observed data point -- Coma -- is within the space of NCC halos of TNG-Cluster. With this comparison we conclude, at face value only, that the $z=0$ magnetic field strengths that result in TNG-Cluster due to amplification across cosmic time are reasonable with respect to observational constraints.

\begin{figure}
    \centering
    \hspace*{-1cm}\includegraphics[width=0.46\textwidth]{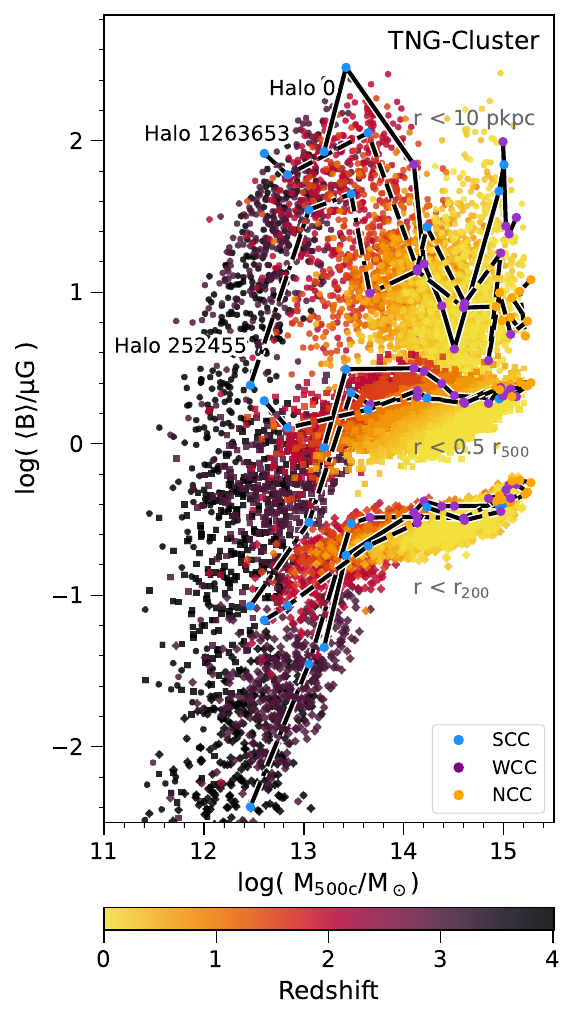}
    \caption{Redshift evolution of magnetic field strength in TNG-Cluster up to $z=4$. We simultaneously show the magnetic field strength within 10\,pkpc (circles), within $0.5\rvir$ (squares), and within $\rtc$ (diamonds) as a function of halo mass $\mvir$. The black lines show the evolution of three individual representative halos for all three apertures. Along these trajectories, blue, purple, and markers indicate the CC state based on central entropy. We see that cluster magnetic fields amplify rapidly at all scales until $z \sim 2$, after which the behavior becomes more complex and depends on distance and CC state.}\label{fig:BvsM500vsZ}
\end{figure}

\subsection{Time evolution of magnetic fields}
In Fig.~\ref{fig:BvsM500vsZ} we show the redshift evolution of the magnetic field strength for all halos in TNG-Cluster at $z=0$ as well as their main progenitors at earlier redshifts up to $z=4$. Since magnetic field strength is a strong function of cluster-centric distance, we show simultaneously three different apertures: within 10 pkpc (top point cloud; circles), within $0.5 r_{\rm 500c}$ (middle; squares), and within $r_{\rm 200}$ (bottom set of points; diamonds).

Magnetic fields in cluster progenitor cores amplify rapidly at early times, and reaches peak values of $\sim 10^3 \mu\rm{G}$ at $z \sim 2-3$. It then declines rapidly towards $z = 0$, settling at characteristic values of $\sim 10^1 \mu\rm{G}$.\footnote{This behavior is qualitatively the same if we instead consider an aperture of 10\,ckpc (not shown).} This sharp decline reflects the onset of strong AGN feedback in these protoclusters, and the resulting impact it has on the gas distribution in their centers. In particular, the SMBH kinetic mode is effective at evacuating gas via strong galactic-scale outflows \citep{nelson2019b}. This causes a non-monotonic dip in halo-scale gas fractions \citep{pillepich2018}, coupling SMBH mass and energetics to gas content \citep{davies2020}, as reflected in the expulsion of gas beyond the halo \citep{ayromlou2023}. This begins the process of galaxy quenching \citep{weinberger2018,terrazas2020}. In \citet{nelson2018}, we showed that a consequence of this gas redistribution is a dichotomy in the magnetic field strength at fixed halo mass, whereby blue star-forming galaxies have much higher magnetic fields than their red, quenched counterparts. At the characteristic halo mass of $\sim 10^{12} \msun$ where central galaxies begin to quench, the result is a drop in central magnetic field strength, as we see here in the high-redshift progenitors \citep[see also][for a recent analysis of magnetic field growth in two cluster zooms simulated with the TNG model]{tevlin2024}.

\begin{figure*}
    \centering
    \hspace*{-1cm}\includegraphics[width=0.7\textwidth]{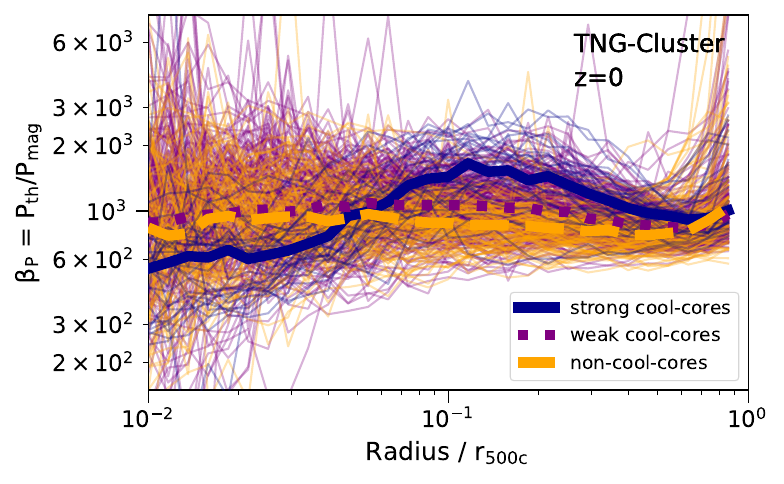}
    \includegraphics[width=0.49\textwidth]{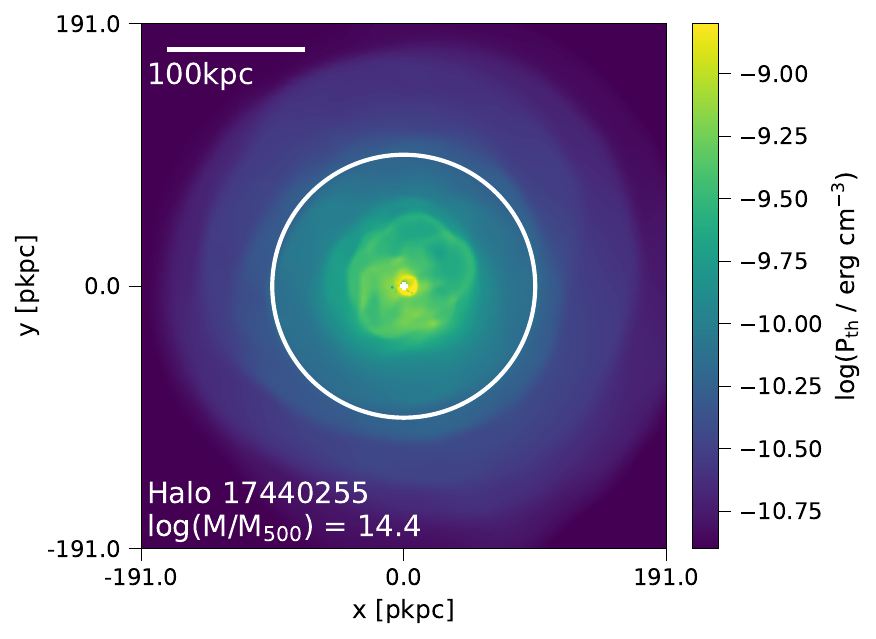}
    \includegraphics[width=0.49\textwidth]{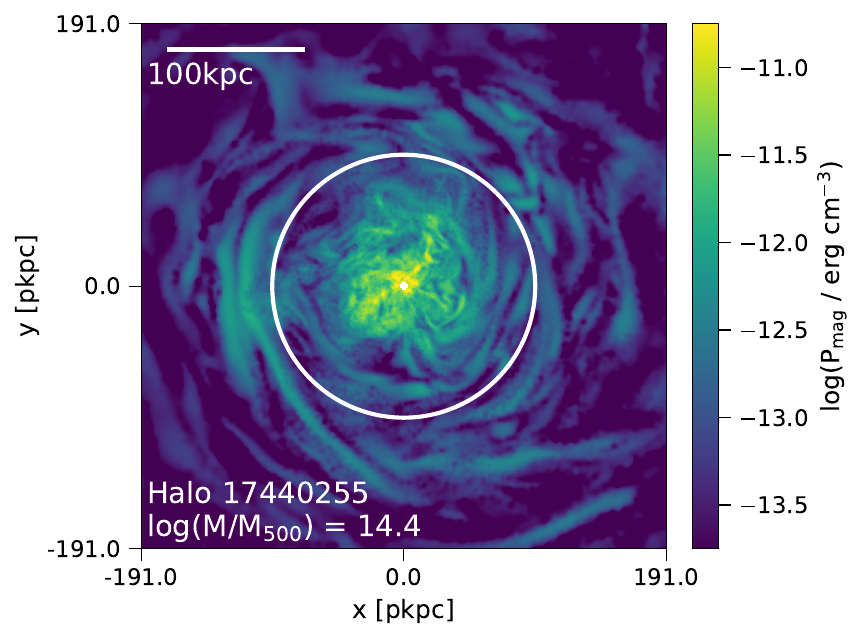}
    \caption{\textbf{Main panel:} radial profiles of the plasma beta parameter $\beta_{\rm P}$ at $z=0$ for all TNG-Cluster halos, colored according to CC state. Thin lines show individual clusters, while the three thick lines show median stacks of SCC, WCC, and NCC clusters. \textbf{Lower panels:} maps of thermal pressure (left) and magnetic pressure (right) of one individual SCC halo ans an example. Both maps show the central region with $r<0.2\rvir$ for a slice with depth $15$\,kpc. The overall pressure budget is dominated by the thermal pressure for both CCs and NCCs. $\beta_{\rm P}$ profiles for WCCs and NCCs are flat with average values of $\beta_{\rm P}\sim 900$. SCCs have a small dip in the core and a excess at $0.1 \rvir$. However, these trends are rather weak.}\label{fig:profileBetaP}
\end{figure*}

In TNG-like models we know that magnetic field amplification proceeds rapidly towards saturation, and reaches equipartition with turbulent energy densities for sufficiently massive galaxies \citep{pakmor2024}. This happens early for massive objects. 

The black lines in Fig.~\ref{fig:BvsM500vsZ} show the evolution of magnetic field strength within the specified aperture for three
representative halos in TNG-Cluster with a mass of $\mvir = 1.26 \times 10^{15}\msun$ at $z=0$. These evolutionary tracks highlight the diversity in the histories of cluster cores and connects to the cool-core (CC) state of clusters. As shown in \cite{lehle2025}, cluster cores are predominantly in the CC state at high redshifts and evolve toward the NCC regime over time. Consequently, a decrease in central magnetic field strength is expected with this evolution, since CC status as well as magnetic field strength depend on the gas density. We illustrate the link between the CC state of a cluster and its central magnetic field strength using colored dots along the evolutionary tracks of these halos. Blue dots indicate a SCC state, purple dots indicate WCC states, while orange dots represent NCC states.

\begin{figure*}
    \centering
    \includegraphics[width=0.9\textwidth]{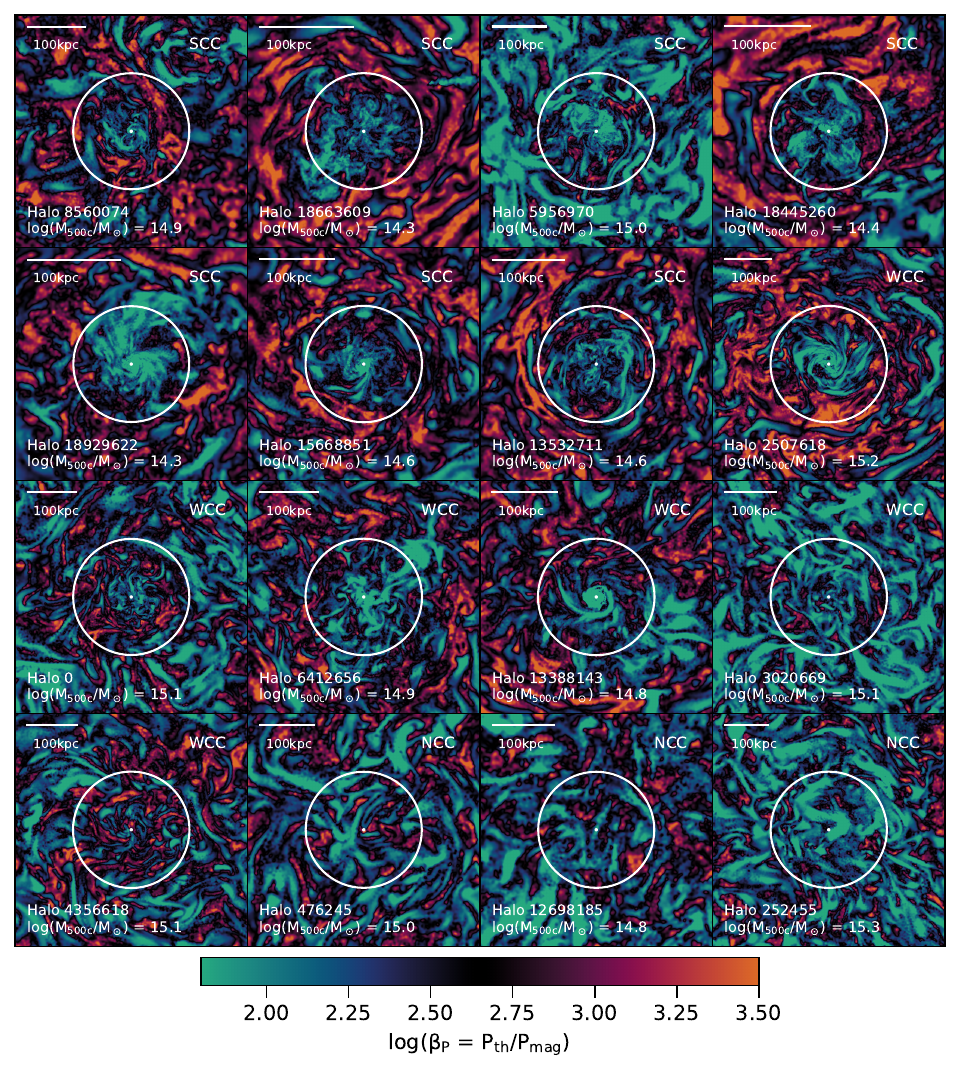}
    \caption{Gallery of the plasma beta parameter $\betaP$ at $z=0$ for the same sixteen TNG-Cluster halos and the same setup as in Fig.~\ref{fig:galleryB}. Each panel shows the central region and extends $0.2\rvir$ from side-to-side, while the white circles indicate $0.1\rvir$. As the panels are ordered by ascending central entropy $K_0$, we also see that in some of the SCCs the decrease of $\betaP$ toward the core is well visible. We also find lower beta for more massive clusters, in agreement with \protect\cite{marinacci2018}.}\label{fig:galleryBetaP}
\end{figure*}

\subsection{Plasma beta parameter}

To characterize the dynamical importance of magnetic fields in clusters, as they transition between CC and NCC states, we turn to the plasma beta parameter $\betaP$. Fig.~\ref{fig:profileBetaP} presents the radial profiles of $\betaP$ for all TNG-Cluster halos, with individual profiles color-coded by CC status. The profiles exhibit considerable scatter on a halo-by-halo basis, with the largest scatter in the core ($\log \betaP \sim 1 - 4 $). Toward larger radii the scatter continuously decreases, but also gains a tail towards higher $\betaP$ values. Despite the scatter among individual halos, the median profiles show distinct trends. WCCs and NCCs halos have flat profiles with typical values $\betaP \sim 10^3$. In contrast, SCCs exhibit a different behavior: $\betaP$ increases from $\rvir$ to $0.1\rvir$, reaching values of approximately $2000$, before dropping sharply toward the core to about $500$ at $0.01\rvir$.

The trends in the median profiles cannot be explained by simple theoretical arguments of flux freezing. Assuming isotropic magnetic fields $\betaP$ would scale with density and temperature as $\betaP \propto T \rho^{1/3}$. Adopting the median temperature and density profiles from \cite{lehle2024} for the TNG-Cluster sample, we find that the median profile for NCCs would indeed be flat and in agreement with Fig.~\ref{fig:profileBetaP}. However, the profiles for WCCs and SCCs would monotonically decrease with decreasing radius and would have a plateau in the center. The decrease would be smaller, and plateau at larger radii, for WCCs. This implies that the rise of $\betaP$ at $0.1\rvir$ in SCCs cannot be explained using these simple arguments. In particular, there must be additional processes shaping the magnetic field strength in these clusters. Importantly, as we show below, magnetic fields in CCs are not isotropic, thus simple flux freezing arguments assuming isotropy are not applicable. 

The central drop in $\betaP$ observed in SCCs is driven by the strong increase in magnetic field strength, as shown in Fig.~\ref{fig:profileBetaP}, which dominates the change in thermal pressure. The rise in $\betaP$ from $\rvir$ to $0.1 \rvir$ predominantly occurs because the thermal pressure increases more significantly than the magnetic pressure. In contrast, the median profiles in WCCs and NCCs remain flat, as the increases in thermal and magnetic pressure are of comparable magnitude across the entire range.

The lower panels of Fig.~\ref{fig:profileBetaP} show maps of thermal and magnetic pressure for the central $0.2\rvir$ of a single halo, for illustration. The thermal pressure map is rather smooth across the halo, although it shows a rich phenomenology of ripple-like features towards the cluster core, indicating shocks, sound waves, and under-dense cavities driven by AGN feedback \citep{truong2024,prunier2025,prunier2025a}. In contrast, the magnetic pressure map has much more significant small-scale variation. As a result, the spatially resolved map of $\betaP$ varies on small scales and the dynamical importance of magnetic fields is a complex function of local properties. Nevertheless, thermal pressure always dominates globally over magnetic pressure.

Fig.~\ref{fig:galleryBetaP} visualizes the same sixteen halos with the same configuration as in Fig.~\ref{fig:galleryB}, but this time showing plasma beta. The $\betaP$ values range from 20 to 4000, and more massive halos generally have lower $\betaP$. For some cool-core halos, the distinctive `excess' feature seen in the SCC median profile of Fig.~\ref{fig:profileBetaP} is also visible here. In particular, $\betaP$ is frequently non-monotonic and peaks at $\sim 0.1\rvir$, decreasing towards the core as well as the outskirts. The halos with the strongest trend in $\betaP$ are also the clusters with the strongest increase of magnetic field strength in Fig.~\ref{fig:galleryB}. 

We note that the profiles and maps of $\betaP$ include all gas, regardless of its phase, and in particular, we take volume-averaged values. As a result, they are dominated by the hot, volume-filling phase in clusters. If we instead consider only cold gas ($T<10^{5}$\,K) and re-compute $\betaP$, the resulting values drop significantly to $10^{-7} - 10^{-3}$ (not shown). This indicates that magnetic fields can be dynamically important in cold gas clouds where $\betaP \ll 1$. This is a generic effect seen in simulations of isolated ellipticals and clusters \citep{Wang2020,Wang2021} and in TNG50 \citep{nelson2020}. However, magnetic fields are not dynamically important in the volume-filling hot ICM.

\subsection{Magnetic field anisotropy}\label{sec_morphology}

\begin{figure*}
    \centering
    \hspace*{1.2cm}\includegraphics[width=0.85\textwidth]{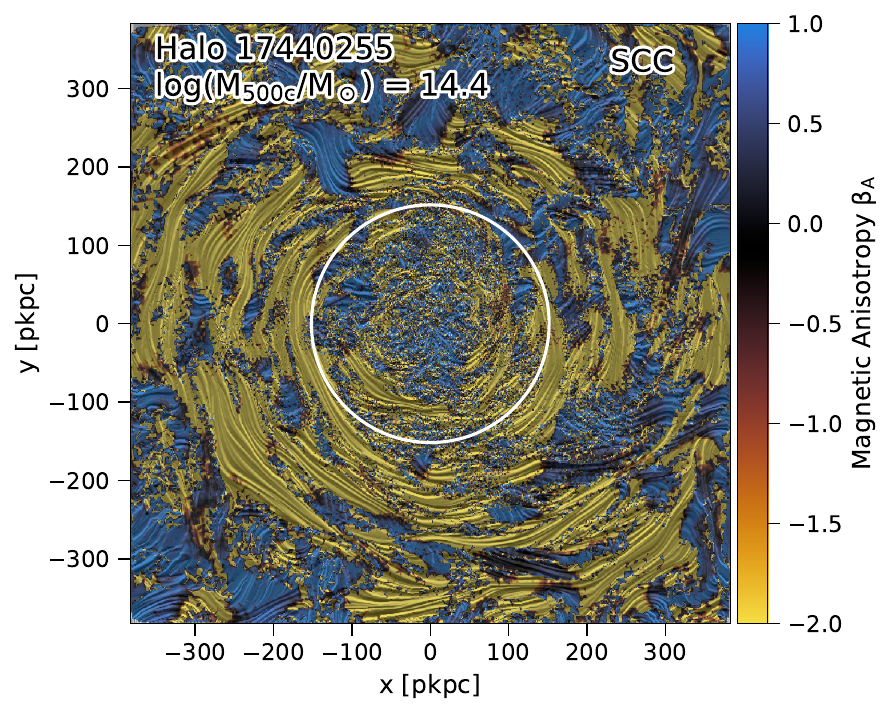} \\
    \hspace*{-1.2cm}\includegraphics[width=0.7\textwidth]{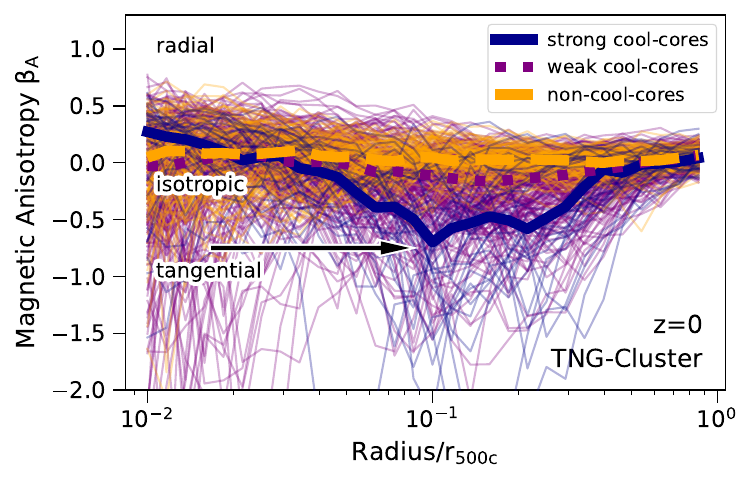}
    \caption{The topology of magnetic fields can be preferentially tangential for cool-core clusters. \textbf{Top panel:} Visualization of magnetic fields in a single TNG-Cluster halo at $z=0$ (Halo 17440255) showing the central $40\%$ of $\rvir$. The background color shows the magnetic anisotropy $\betaA$ in the $xy$-plane. Negative values (yellow) correspond to tangential magnetic fields and positive values (blue) correspond to radially oriented magnetic fields. Overplotted in gray are the magnetic field directions, which appear as relief-like structures in the image. The white circle marks the radius at which the magnetic anisotropy profile (bottom panel) has its minimum. \textbf{Bottom panel:} Radial profiles of magnetic field anisotropy parameter $\betaA$ at $z=0$ for all halos in TNG-Cluster. Thin lines show individual halos, colored by CC state. The thick lines show the median profiles for SCCs (blue), WCCs (purple) and NCCs (orange). WCCs and NCCs show a flat profile at $\betaA\sim0$, while SCCs exhibit a dip at $\sim 0.1\rvir$ with $\betaA <0$. This shows that WCCs and NCCs have primarily isotropic magnetic fields, while SCCs have preferably tangentially oriented magnetic fields in the vicinity of this radius (as indicated by black arrow).}\label{fig:betaA}
\end{figure*}

We see that the magnetic fields in the cores of clusters can exhibit diverse morphologies. To quantify the overall topology and orientation of magnetic field lines, we compute magnetic anisotropy profiles $\betaA(r)$ for our TNG-Cluster halos (as defined in Sec.~\ref{eq:betaA}). As $\betaA$ measures the ratio of the tangential versus radial components of the magnetic field, $\betaA<0$ for tangentially dominated fields, $\betaA>0$ for radially dominated fields, and $\betaA\sim 0$ for fields with no preferential orientation (i.e. isotropic).

Fig.~\ref{fig:betaA} shows the magnetic anisotropy of TNG-Cluster halos, and is the main finding of our work. It visualizes the spatial structure of anistropy for a single representative SCC halo (top panel), and shows radial $\betaA$ profiles for all halos in TNG-Cluster colored by CC status (bottom panel).

The visualization shows $\betaA$ within a thin slice of the central $0.4\rvir$ of the cluster, oriented in an arbitrary direction. The background color encodes the magnetic anisotropy in the $xy$-plane: yellow indicates a tangential orientation of magnetic fields, while blue shows a radial orientation. In gray we show the local directions of magnetic field vectors in the same plane, which appear as relief-like structures in the image. We see a striking feature: while magnetic fields in the core tend to be tangled with isotropic and/or radial orientation, magnetic fields become preferentially tangentially oriented (yellow) at a characteristic radius, indicated by the white circle.\footnote{The white circle is located at the radius at which the magnetic anisotropy profile of this single halo has the minimum.} It is also clear that this feature is made up of patches of tangential fields, rather than a continuous tangential layer wrapping around the core.

Fig.~\ref{fig:betaA} (bottom panel) reveals that most clusters have isotropic magnetic field orientations, with profiles centered around $\betaA\sim 0$. There is, however, significant scatter among individual profiles. The scatter is smallest at the largest radii, where the magnetic fields are consistent with an isotropic orientation. Moving toward the center, most profiles remain compatible with isotropy, but around $\sim0.1\rvir$, the scatter increases significantly, with many clusters showing negative $\betaA$ values. Towards the core, the scatter decreases, but becomes largest at the center. 

Most striking, we see a clear connection between magnetic field anisotropy and cluster cool-core status (bottom panel). The three thick lines show median profiles when we split the cluster population between our three cool-core states.\footnote{There are no relevant differences between mean and median profiles.} The magnetic fields in WCCs (purple) and NCCs (orange) are isotropic across all radii. In contrast, the median profile for SCCs has a prominent dip at $\sim 0.1 \rvir$, indicating preferentially tangentially oriented magnetic fields. This feature is absent in NCC clusters.

Despite this clear signature in the median SCC profile, there is additional variation and diversity between individual halos. For instance, there are several WCCs that also show strong dips, reflecting the rather arbitrary threshold that separates CCs from NCCs \citep{lehle2024}. Furthermore, the dip in individual SCC profiles varies in location, with the most tangential fields (lowest $\betaA$ values) occurring across a radial range of $r\sim 0.05- 0.2\rvir$.

\begin{figure*}
    \centering
    \includegraphics[width=0.49\textwidth]{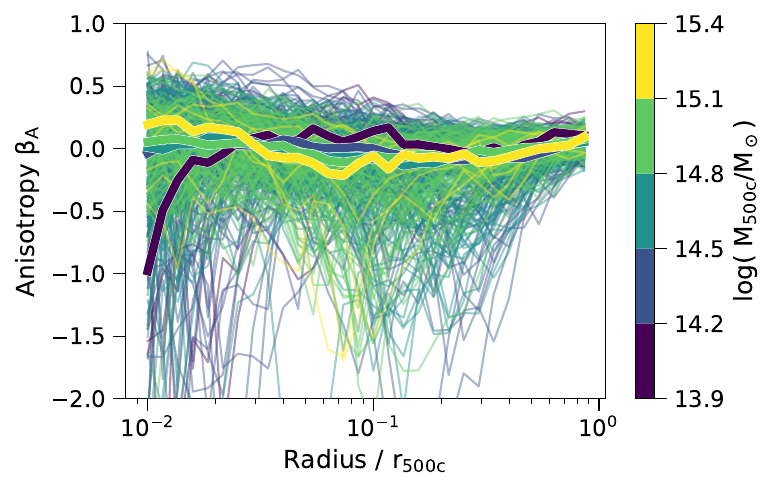}
    \includegraphics[width=0.49\textwidth]{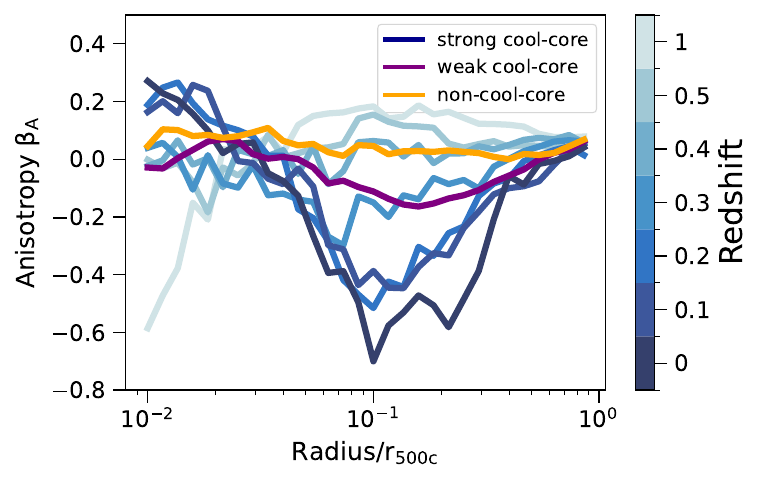}
    \caption{\textbf{Left:} Mass trends of the anisotropy profiles for all halos in TNG-Cluster. The thin lines show individual halos, while the thick lines show medians in mass bins. There are only weak trends in the average magnetic anisotropy with cluster mass.  \textbf{Right:} Median redshift trends of the anisotropy profiles for all SCCs in TNG-Cluster; the redshift evolution for WCCs and NCCs is not shown, as it is consistent with a flat radial profile for all redshifts. In SCCs, the population wide dip in $\betaA$ decreases toward higher redshifts, until $z=0.4$, where the profiles become flat and consistent with the WCC and NCC profiles.}\label{fig:betaA_mass_z}
\end{figure*}

In contrast to this clear difference between CC and NCC status, Fig.~\ref{fig:betaA_mass_z} shows that there is only a weak trend in the average with cluster mass (left panel). Across different mass bins the profiles are all compatible with isotropic magnetic field orientations. Halos in the smallest mass bin always have $\betaA <0$ in the core. Most importantly, clusters with highly negative values at $\sim 0.1\rvir$ are found among all mass bins. If we only consider SCCs, the dip for more massive SCCs is less deep and can be found at smaller radii than for less massive SCCs (not shown).

In order to count the fraction of clusters that exhibit a dip in $\betaA$ at $z=0$, we consider the $\betaA$ profile outside the core ($r > 0.03$), and identify a strong dip by requiring $\beta_{\text{min}} < -0.5$ ($\beta_{\text{min}} < -0.75$). According to this definition, 136 (83) clusters in our $z=0$ TNG-Cluster sample have prominent tangential magnetic field features. Breaking this down by cluster type, we find that out of 31 SCCs, 28 (24) meet the criteria for a dip. Among 215 WCCs, 102 (59) have a dip and for the 106 NCCs in our sample, only 6 (0) exhibit a dip.

Finally, we examine the redshift evolution to determine whether the pronounced tangential magnetic fields in SCCs persist at higher redshifts. Fig.~\ref{fig:betaA_mass_z} (right panel) shows the median redshift evolution for all SCC halos in TNG-Cluster. We do not include WCCs and NCCs in this figure, as their profiles remain consistent with isotropic magnetic fields across all redshifts. For SCCs, the dip in $\betaA$ decreases with increasing redshift, and by $z=0.4$, the profile is flat.\footnote{We track the evolution of SCC by reclassifying the CC status at each redshift. If we instead identify SCCs at earlier times as the direct progenitors of $z=0$ SCCs, the overall trend remains the same. That is, the dip in $\betaA$ diminishes at higher redshifts.}

Could the redshift trend result from an overall decline in the mean halo mass of the sample at higher redshifts? We consider Fig.~\ref{fig:betaA_mass_z} (left panel), showing that the dip in $\betaA$ is strongest in less massive SCCs and gradually weakens with increasing cluster mass. This implies that, if the redshift evolution of the population-wide trend was driven by mass evolution, the dip should remain present at higher redshifts and might even become more prominent. This is not the case, suggesting a difference in the driving mechanism and/or physical state of the ICM needed to support such tangential features, as we discuss further below (Sec.~\ref{sec_discussion}).

\begin{figure*}
    \centering
    \includegraphics[width=0.38\textwidth]{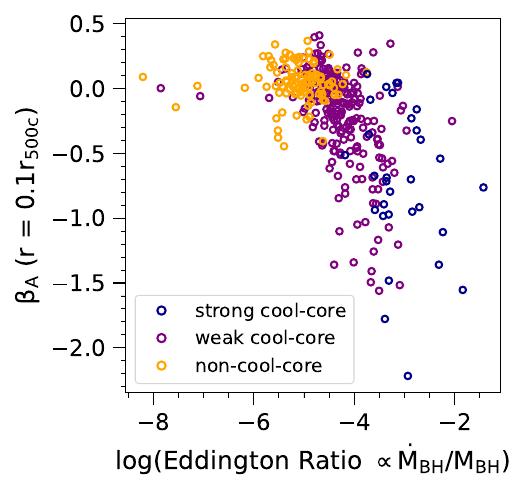}
    \includegraphics[width=0.38\textwidth]{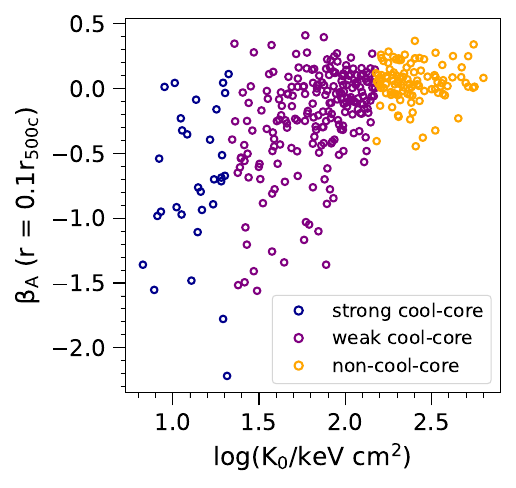}\\
    \includegraphics[width=0.38\textwidth]{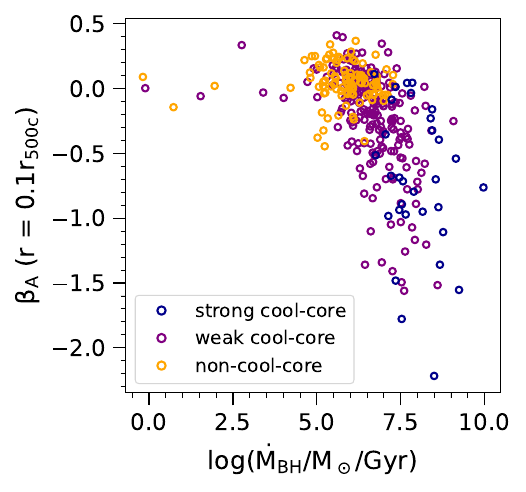}
    \includegraphics[width=0.38\textwidth]{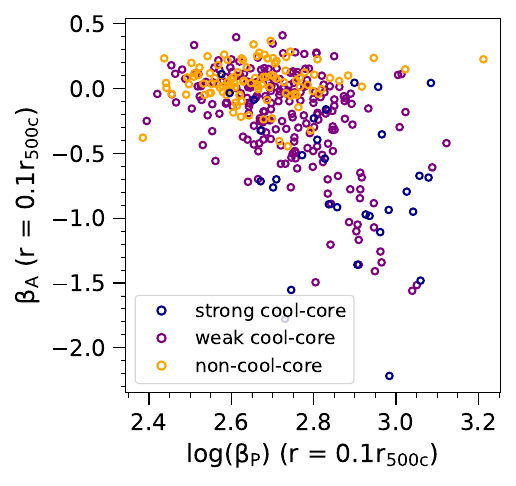}
    \caption{Correlation of the magnetic anisotropy $\betaA$ measured at $0.1\rvir$ with various halo properties for the $z=0$ TNG-Cluster sample. The markers are colored according to CC state. $\betaA(r=0.1\rvir)$ is calculated from a shell with thickness $0.01\rvir$. Low values of $\betaA$, corresponding to tangential magnetic field topologies, are found in clusters with (i) high Eddington ratios (upper left),  (ii) low central entropy (upper right), (iii) high SMBH accretion rates (lower left), and (iv) high plasma beta $\betaP(r=0.1\rvir)$ (lower right).}\label{fig:betaA0p1_correl}
\end{figure*}

\subsection{Correlations of magnetic field anisotropy with halo, galaxy, and SMBH properties}
 
To probe the tangential orientation of the magnetic fields in some clusters, and as a first step toward identifying the factor(s) driving this morphology, we examine relationships between $\betaA(r=0.1\rvir)$ and halo, galaxy, and SMBH properties. We measure $\beta$ at this radius as it is the location of the minimum in the median SCC profile.

Fig.~\ref{fig:betaA0p1_correl} shows four correlations that we identify as the strongest and most informative. First, we highlight the connection between magnetic field anisotropy and AGN activity (upper left panel), quantified in terms of the SMBH Eddington ratio
\begin{equation}
   \eta_{\rm Edd} =  \frac{4\pi G m_p}{\varepsilon_r \sigma_T c}\frac{\dot{M}_\mathrm{BH}}{M_\mathrm{BH}}\propto \frac{\dot{M}_\mathrm{BH}}{M_\mathrm{BH}}\, , 
\end{equation}
of the most massive central SMBH. Here, $G$ is the gravitational constant, $m_p$ the proton mass, $\varepsilon_r = 0.2$ the black hole radiative efficiency, $\sigma_T$ the Thompson cross-section, $c$ the speed of light, $M_\mathrm{BH}$ the black hole mass, and $\dot{M}_\mathrm{BH}$ the instantaneous black hole accretion rate.

The scatter plot reveals a clear correlation: clusters with higher Eddington ratio SMBHs have lower and more negative $\betaA$ values. Furthermore, SCCs (marked in blue) are associated with lower values of $\betaA$, indicative of tangentially oriented magnetic fields and higher Eddington ratios reflecting strong instantaneous AGN activity. In contrast, NCCs (in orange) show distributions closer to isotropy or positive $\betaA$ and lower Eddington ratios. WCCs (in purple) exhibit a wide range of values in both $\betaA$ and Eddington ratios.\footnote{Considering the uncertainties in classifying SCCs, WCCs, and NCCs, the WCC population can be viewed as polluted by NCCs on one end of the parameter space and with SCCs on the other.}

Fig.~\ref{fig:betaA0p1_correl} (upper right) shows the relation between $\betaA$ and the central entropy $K_0$. SCCs, by definition with low central entropy, tend to exhibit more negative $\betaA$. NCCs have the highest central entropy and $\betaA\sim 0$. We also show the accretion rate $\dot{M}_\mathrm{BH}$ of the most massive SMBH (lower left), with a similar trend: higher accretion rates correlate with more negative $\betaA$. Finally, we explore the plasma beta parameter $\betaP$ at $0.1\rvir$ (shell thickness $0.03\rvir$ (lower right), indicating that higher $\betaP$ values are also linked to more negative $\betaA$. However, this correlation is rather weak and the separation between SCCs and NCCs is only moderate. 

We also examine other properties including halo mass, outflow velocities, velocity dispersion, time since last merger, and magnetic field strength (not shown). These exhibit weak and/or no clear trends with $\betaA$. Correlations with weak trends are found with outflow velocities ($v_{\rm out}\, \searrow \betaA \searrow$) and the time since last merger ($t_{\rm last \, merger} \nearrow \, \betaA \searrow$)

The observed trends with instantaneous feedback-related properties (Eddington ratio, $\dot{M}_{\rm BH}$, $v_{\rm out}$) are suggestive and imply that AGN activity may play a significant role in shaping the magnetic field morphology. However, it is also possible that these correlations might be explained by a common underlying cause, rather than a direct impact of feedback on $\betaA<0$. Below we discuss whether AGN feedback is a possible driver of the tangential magnetic fields (Sec.~\ref{sec_discussion}). 

\subsection{Velocity anisotropy}
\label{velocity_anisotropy}

\begin{figure}
    \centering
    \includegraphics[width=0.48\textwidth]{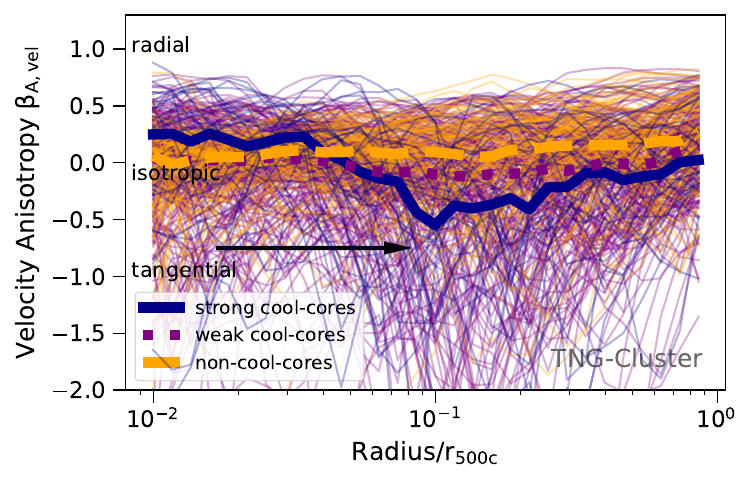}
    \includegraphics[width=0.38\textwidth]{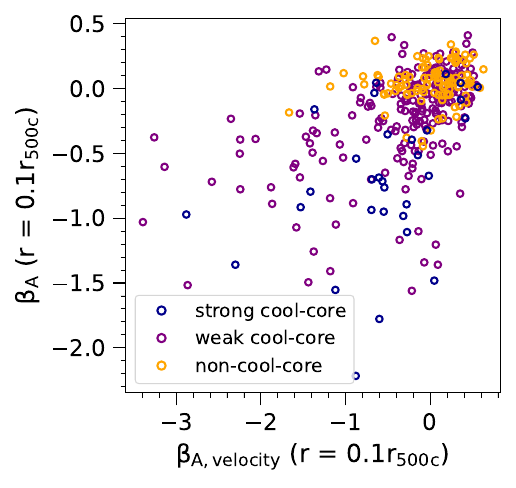}
    \caption{\textbf{Top:} Radial profiles of velocity anisotropy, where thin lines show individual halos from TNG-Cluster at $z=0$, and thick lines show median stacks. The median profiles for WCCs and NCCs are compatible with an isotropic velocity fields. There is a somewhat population wide tangential bias in SCCs, albeit smaller compared to the magnetic anisotropy. Importantly, we find among all cluster types many individual examples of halos with dips at all radii. \textbf{Bottom:} the correlation between magnetic and velocity anisotropy parameters, both measured at the same $r = 0.1 r_{\rm 500c}$.}
    \label{fig:betaAvel}
\end{figure}

Is the magnetic anisotropy linked to the motion of the ICM? To address this question
we compute the anisotropy of the velocity field by adopting Eq.~\ref{eq:betaA} where we simply replace the magnetic field by the velocity field.

Fig.~\ref{fig:betaAvel} shows the result: radial profiles of velocity anisotropy (top panel) and the relationship between magnetic and velocity anisotropies (bottom panel). In the former, thin lines show individual halos, colored by cooling status, while thick lines show median profiles. As in the case of magnetic anisotropy, the median velocity anisotropy for WCCs and NCCs are flat and compatible with an isotropic configuration. For SCC clusters, there is a population-wide tangential bias in the velocity field. However, this extends to smaller deviations from isotropy than in the case of magnetic fields. Importantly, there are many individual halos among SCCs, WCCs, and NCCs, that have tangentially biased velocities fields at all radii. The population-wide difference between the SCCs and the NCCs velocity anisotropy is significantly smaller than the difference in the magnetic anisotropy between these cluster categories.

The bottom panel shows that magnetic anisotropy has at least a weak positive trend with velocity anistropy. In particular, more isotropic magnetic fields go hand in hand with more isotropic velocity fields. While magnetic fields are advected with the fluid under the flux-freezing condition, the presence of directional bias in the magnetic field (e.g., anisotropy or preferential alignment) does not inherently necessitate a simultaneous anisotropic velocity field. Instead, anisotropic velocity fields -- such as shear flows or stratified motions -- can imprint directional biases onto the magnetic field through inductive processes. Once established, the magnetic field may retain a cumulative memory of past fluid motions even after the original velocity structures responsible for its organization have decayed. This memory arises from the high electrical conductivity of astrophysical plasmas (i.e., high magnetic Reynolds number), which preserves the magnetic topology against diffusion. Thus, the magnetic field can serve as a historical record of prior dynamical conditions, reflecting the integrated effects of anisotropic gas motions.

Our simulations employ ideal MHD, and thus assume flux-freezing conditions. The magnetic fields can be affected by subsequent stirring motions; however, the extent to which this occurs depends on the specific details of the flow. Consequently, some degree of magnetic field ordering can survive such stirring, especially when the Froude number is small. Overall, it is not necessarily the case that the anisotropy of the magnetic field should be mirrored in the velocity field when measurements are averaged over extended periods of time.


\section{Discussion} \label{sec_discussion}

We have identified a population-wide preferential tangential orientation of magnetic fields in SCC clusters at a characteristic distance of $\sim 0.1 \rvir$. In this section, we explore and discuss potential causes that might explain the origin of these magnetic field structures. The following effects are known to shape magnetic fields in the ICM: 

\begin{itemize}
    \item Heat flux-driven buoyancy instability (HBI) - HBI arises in the presence of anisotropic thermal conduction when the temperature decreases in the direction of gravity, as seen in CCs. Since TNG-Cluster does not include anisotropic transport processes, the HBI cannot occur. In addition, it is unclear whether HBIs can affect magnetic fields in the ICM, because (i) for realistic magnetic field strengths in clusters, magnetic tension can suppress the HBIs \citep{drake2021}, (ii) field-aligned thermal conduction may be suppressed by the whisteler instability \citep{drake2021}, and (iii) even weak turbulent motions can suppress the HBI \citep{parrish2010,ruszkowski2010}.
    \item (Sec.~\ref{sec_disc_AGN}) Outflows from SMBH feedback can produce tangentially biased magnetic fields via the formation of AGN-driven shocks or via draping of field lines around rising bubbles.
    \item (Sec.~\ref{sec_disc_merger}) Mergers can induce large-scale motions, such as sloshing, or create shocks, both of which can influence the structure of magnetic fields.
    \item (Sec.~\ref{sec_disc_gwaves}) Under certain conditions, gravity waves can be excited and trapped within the ICM, potentially influencing the orientation of magnetic fields.
\end{itemize}

\subsection{AGN feedback - bubbles and shocks}\label{sec_disc_AGN}

SMBHs can generate powerful outflows through feedback processes that significantly impact the cores of galaxy clusters, as suggested by observations \citep{mcnamara2009, nulsen2005} as well as TNG model simulations \citep{pillepich2021, nelson2019b}. We see a strong correlation between black hole properties and the magnetic field orientation at $0.1 \rvir$ for halos in TNG-Cluster (Fig.~\ref{fig:betaA0p1_correl}). We therefore consider the possibility that this correlation reflects a causal relationship. If AGN feedback indeed induces tangentially biased magnetic fields, this could occur through magnetic field draping around rising bubbles or via fast shocks, both resulting from energy injections in the kinetic feedback mode.

AGN activity in SCCs is marked by high accretion rates, high Eddington ratios, and strong outflows in the core. In \citet{lehle2025} we show, by tracing the evolution of a cluster with high time resolution, that individual AGN feedback events drive powerful outflows. To understand their impact on our tangential magnetic field feature, we reconsider this analysis while measuring the outflow velocities at $0.1\rvir$. At this radius, AGN feedback does not significantly move large amounts of gas; only some high-velocity gas tails are present \citep[making such outflows also hard to identify in XRISM-type high energy resolution X-ray spectroscopy;][]{truong2024}.

This same AGN feedback generates bubbles that are observable in the form of X-ray cavities, and these are more common in SCCs than in WCCs or NCCs \citep{prunier2025}. However, X-ray cavities in observations and TNG-Cluster are generally found at smaller radii ($\sim 20-40$\,kpc on average) than where we find the population-wide tangentially biased magnetic fields \citep{prunier2025}. This suggests that the draping of magnetic field lines around bubbles is unlikely to account for the magnetic anisotropy at $\sim 0.1\rvir$. In the individual case shown in Fig.~\ref{fig:betaA}, which exhibits prominent tangential magnetic fields at $\sim 150\,\mathrm{kpc}$, we also find an X-ray cavity. However, the cavity lies well within the region of tangential magnetic field structure. Thus, at least in this case, AGN-driven bubbles do not seem to directly produce the tangential magnetic fields.

However, the shocks generated by AGN feedback can occur at larger distances, reaching typically $\sim 50\,\mathrm{kpc}$, with some extending as far as $120\,\mathrm{kpc}$ (\textcolor{blue}{Prunier et al. in prep}). Furthermore, such shocks are more prominent in SCCs than in NCCs. For these AGN-driven shocks to align post-shock magnetic fields tangentially -- i.e. parallel to the shock plane -- they must be fast shocks \citep{lehmann2016}. Fast shocks are characterized by an Alfvénic Mach number $\mathcal{M}_A > 1$, where
\begin{equation}
\mathcal{M}^2_A = \frac{v^2}{v^2_A} = \frac{1}{2}\gamma \betaP \mathcal{M}^2\, ,
\end{equation}
where $v_{A}$ is the Alfvén speed. The shocks presented in \textcolor{blue}{Prunier et al. (in prep)} have Mach numbers of $1 - 1.8$, with a median value of $1.1$. This implies that all TNG-Cluster shocks are fast shocks, given the values of $\betaP>100$ that we reported. Thus, they would be able to tangentially align the post-shock magnetic fields. However, ICM shocks are essentially non-radiative, and the Rankine-Hugoniot jump conditions at high plasma beta reduce to standard hydrodynamical jump conditions. As a result, small Mach numbers give small density jumps, and correspondingly small changes to the magnetic field component parallel to the shock. This implies that a strong tangential magnetic field bias cannot be caused by weak shocks amplifying magnetic field strengths \citep[see Eqn. 6 of][]{granda-munoz2025}.

Returning to the same individual halo (as shown in Fig.~\ref{fig:betaA}), we identify a fast shock at a distance of approximately $80 - 90\,\mathrm{kpc}$. This radial location is still closer to the cluster center than the tangential magnetic field feature. The mach number of such shocks will rapidly decrease as they propagate and eventually fade into sound waves. The spatial location offsets between the typical fast shock position and the $\beta_A$ dip are therefore somewhat inconclusive.

However, if AGN feedback causes the preferential tangential orientation of magnetic fields, it should also explain the fact that the population-wide tangential bias of SCCs vanishes at higher redshift. At earlier times, halos are less massive with shallower gravitational potential wells. If the dip in $\betaA$ is caused by AGN feedback, it may be expected to shift to larger radii since the effects of AGN feedback could extend farther in a less massive halo \citep[e.g.][and references therein]{oppenheimer2021}. This could explain our finding that the dip in $\betaA$ moves to larger distances at higher redshift, when we consider the progenitors of SCCs. However, as highlighted in the discussion of Fig.~\ref{fig:betaA_mass_z}, the observed redshift evolution across the population cannot be explained by a decline in sample mass at higher redshifts alone. This is because less massive SCCs tend to show a more pronounced dip in magnetic anisotropy at $z=0$, suggesting that the magnetic fields should still remain tangential at higher redshift, whether the dip in $\betaA$ moves to larger radii or not. Finally, we also note that at $z>0$, SCCs still show consistently higher Eddington ratios than NCCs, yet the tangential magnetic field signal ($\betaA<0$) disappears at higher redshifts. 

AGN feedback could explain the tangential magnetic fields at $z=0$ plus isotropic fields at higher redshifts if there was a time lag between AGN activity and the formation of tangential magnetic fields. In this case, the AGN feedback would require a significant amount of time to generate the tangential fields. However, this time lag would need to be long ($\sim 5\, \mathrm{Gyr}$), which is not the case in the TNG model, where AGN activity happens continuously. Furthermore, such a scenario is unlikely given that tangential fields can appear and disappear between consecutive snapshots in many SCCs with tangential magnetic fields at $z=0$ (e.g. see Fig.~\ref{fig:timeEvo}). 

Given these considerations, although there are population-wide differences between SCCs and NCCs in terms of AGN activity, we argue that AGN feedback does not cause the preferential tangential orientation of the magnetic fields in the central region of SCCs. We predominantly expect AGN feedback to shape the magnetic field morphology closer to the center than at $\sim 0.1\rvir$. Once there is a final catalog of shocks from TNG-Cluster, a more quantitative and direct comparison can be done to study the relationship between shocks and magnetic field topology.

\subsection{Mergers - sloshing motions and shocks}\label{sec_disc_merger}

Mergers between halos can induce large-scale motions in the ICM, including sloshing and the generation of powerful shocks. Observational signatures include large-scale radio relics, as also appear in TNG-Cluster halos \citep{lee2024}. The associated shocks have Mach numbers $\mathcal{M}>1.3$, classifying them as fast shocks capable of tangentially aligning post-shock magnetic fields \citep{lehmann2016}. However, radio relics are typically located at much larger radii than $r\sim 0.1\rvir$.

For instance, in halo 3135769, a radio relic is present at $r \sim 0.4\rvir$, while a dip in $\betaA$ is observed at $r\sim 0.1\rvir$. This indicates that the radio relic lies well outside the region of tangential magnetic fields. Although these features are misaligned in radius, it remains possible that the associated shocks induce bulk flows affecting these regions. However, velocity maps of this halo show no such flows. In addition, many of the clusters with a dip in $\betaA$ do not show signs of a shock/radio relic or merger (e.g., Halo 17440255, Fig.~\ref{fig:betaA}). These findings suggest that, although possible in some individual cases, merger shocks cannot explain the population-wide tangential bias in magnetic fields at $r\sim 0.1\rvir$ at $z=0$.

Observationally, \citet{hu2020} infer that magnetic fields align with sloshing fronts in the Perseus cluster, while also tending to drape around feedback-driven bubbles and orbiting substructures. The results of \citet{hu2020} rely on the gradient technique, enabling an indirect inference of magnetic field direction, although the method has been verified in several regimes \citep{hu2025}. Overall, this leads to a preferential orientation of magnetic fields in the azimuthal, i.e., tangential direction, similar to our findings from TNG-Cluster for SCCs as a whole. The qualitative similarity between our population wide tangential bias in SCCs and their findings is compelling and suggests that a larger statistical observational sample will be useful in determining the prevalence and possible origins of unique magnetic field configurations in clusters \citep{choudhury2024}.

Mergers not only generate shocks, but also drive large-scale motions that can align magnetic fields with the flow. If the flow is tangential, the magnetic fields could also become tangential. While the bulk flows driven by mergers could certainly contribute to the population-wide preference for tangential magnetic fields in SCCs, there is no clear reason why this effect would consistently occur at a similar radius. Such events could influence a wide range of radii, particularly larger ones, where dips in $\betaA$ should also be observed. However, as shown in Fig.~\ref{fig:betaA}, there is no strong dip with negative magnetic anisotropy for $r>0.4\rvir$. 

Perhaps the most compelling counterargument is the prominent population-wide difference between SCCs and NCCs. While SCCs show tangentially biased magnetic fields, no NCCs exhibit such behavior. If mergers were the sole driver of these magnetic field configurations, we would expect NCCs -- where mergers may even be more frequent -- to occasionally display tangential fields as well. Since this is not observed, we conclude that merger-driven bulk flows and shocks cannot account for the population-wide differences in magnetic field anisotropy between SCCs and NCCs. 

\subsection{Trapping of gravity waves}\label{sec_disc_gwaves}

\subsubsection{Theoretical considerations}

\begin{figure*}
    \centering
    \includegraphics[width=0.8\textwidth]{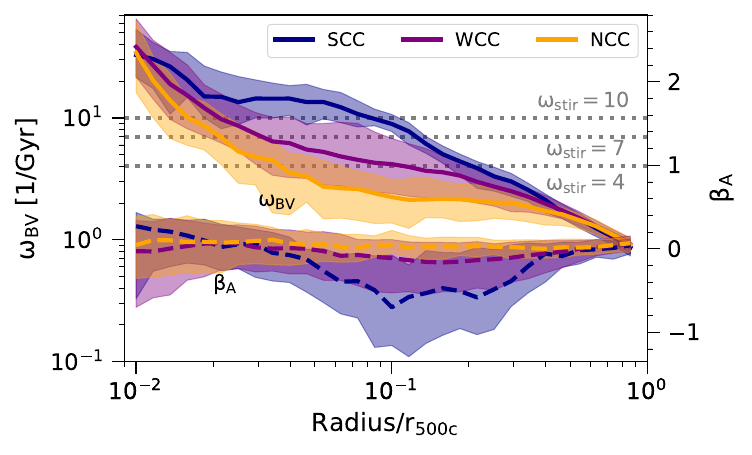}  
    \caption{Comparison of the median Brunt-Väisälä frequency $\omega_{\rm BV}$ radial profiles and the median magnetic anisotropy profiles for all TNG-Cluster halos at $z=0$. The solid lines and left $y$-axis depict $\omega_{\rm BV}$, while the dashed lines and right $y$-axis show the magnetic anisotropy $\betaA$. Each median line and 16th to 84th percentile band is colored according to CC status. The horizontal dotted gray lines indicate exemplary constant stirring frequencies $\omega_{\rm stir}$ in the same units as $\omega_{\rm BV}$, i.e., 1/Gyr. Gravity waves can be exited in regions with $\omega_{\rm stir}<\omega_{\rm BV}$ and are reflected/trapped at a radius where $\omega_{\rm stir} =\omega_{\rm BV}$. The maximum of $\omega_{\rm BV}$, beyond the core, where gravity waves can be most easily trapped, and the minimum in $\betaA$, where magnetic fields show the strongest tangential orientation, appear at similar radii, suggesting that gravity wave trapping can be effective for SCC clusters.}
    \label{fig:wBVvsBetaA}
\end{figure*}

Can preferably tangential magnetic fields in (SCC) clusters be explained by trapped gravity waves? To answer this question, we first review the expectations for gravity wave trapping from linear theory. We consider a hydrodynamical fluid in a stably stratified medium, which is stabilized by entropy gradients. In a medium with a strong entropy gradient, an adiabatically displaced fluid element will oscillate with the Brunt-Väisälä frequency $\omega_{BV}$ 
\begin{equation}\label{eq:wBV}
    \omega_{\rm BV}^2 = \frac{1}{\gamma} \frac{g(r)}{r} \frac{\mathrm{d}ln K(r)}{\mathrm{d}ln(r)}
\end{equation}
around the equilibrium position, with gravity being the restoring force. Following the approach of \cite{balbus1990} of perturbing, linearizing and then applying a WKBJ approach, one obtains a dispersion relation for gravity waves or g-modes\footnote{There is also a solution for p-modes, but as p-waves quickly propagate out through the stratified medium, we will not consider their contribution here.}
\begin{equation}
    \omega^2 = \omega_{BV}^2 \frac{k^2_\perp}{k^2}\, , \label{eq:wBV_disprel}
\end{equation}
where $k^2 = k_r^2 +k^2_\perp$ where $k_r$ is the radial component of the wave vector and $k_\perp$ the tangential component of the wave vector. This dispersion relation implies that gravity waves can be excited for $\omega<\omega_{BV}$, as only then does the condition lead to a real solution. Intuitively, for $\omega<\omega_{BV}$ a perturbation can propagate because the gas has time to react to the perturbation. On the other hand, for $\omega>\omega_{BV}$, there are only imaginary solutions and perturbations are short-lived. 

G-modes have a maximum possible frequency of $\omega_{\rm max} = \omega_{BV}$, at this frequency $k_\perp = |\vec{k}|$ and $k_r = 0$, implying that at the radius where this condition holds true, the waves are purely tangential. If the Brunt-Väisälä frequency is a decreasing function with radius, there is a radius for each frequency, where the g-modes will be reflected. This implies g-modes driven with $\omega<\omega_{BV}$ can be excited and will be trapped, reflected and focused inside the radius at which $\omega=\omega_{BV}$, and the group velocity of the gravity waves is inward \citep{balbus1990}. Accordingly, one can think of this trapping and reflection as similar to electromagnetic waves being reflected by a plasma when the frequency of the wave equals the plasma frequency.  

Gravity waves have been studied in the context of stellar pulsation \citep{cox1980}, galaxy clusters \citep{lufkin1995}, and the Earth’s atmosphere and oceans \citep{riley2000}. In our case, the most relevant aspect is that the g-modes can be tangential. 

\cite{ruszkowski2010} discuss a simple order-of-magnitude argument why this should be the case, and we summarize it here. Let us assume a simple steady-state, and incompressible atmosphere. In such a situation, the continuity equation is $ \vec{v}\cdot\vec{\nabla}\rho = 0$. We split the density $\rho$ in a background and fluctuation contribution $ \rho = \rho_0+ \tilde{\rho}$ and assume that the background density varies only in radial direction. The continuity equation is then $\vec{v}\cdot\vec{\nabla}\tilde{\rho} + v_r (\mathrm d\rho_0 / \mathrm d r) = 0$. Introducing characteristic scales $|\vec{v}|\sim U$, $v_r \sim W$, and length $\sim L$ gives an order-of-magnitude estimation for the density fluctuations
\begin{equation}\label{eq:scalesConti}
    \tilde{\rho}\sim \frac{WL}{U}\left|\frac{\mathrm d \rho_0}{\mathrm d r}\right|\, .
\end{equation}
To find an expression for $\tilde{\rho}$ one can consider the evolution equation for vorticity $\vec{\Omega} \equiv \vec{\nabla} \times \vec{v} $ and again perform a linear perturbation analysis \citep{lufkin1995}. For g-waves one obtains 
\begin{equation}
    \frac{\partial \tilde{\vec{\Omega}}}{\partial t} = i \frac{{\tilde{\rho}}}{\rho} (\vec{k}\times \vec{g})\, .\label{eq:vorticity_generation}
\end{equation}
Applying the same characteristic scale estimation, one obtains
\begin{equation}\label{eq:scalesVorticity}
    \tilde{\rho}\sim \frac{\rho_0 U^2}{g L}\, .
\end{equation}
Combining Eq.~\ref{eq:scalesConti} and Eq.~\ref{eq:scalesVorticity} yields
\begin{equation}
    \frac{W}{U} \sim \frac{\rho_0 U^2}{gL^2 |\mathrm{d}\rho_0/\mathrm{d}r|} \sim \left(\frac{U}{\omega_{BV}L}\right)^2\sim \mathit{Fr}^2\, ,\label{eq:W/U=Fr}
\end{equation}
where $\mathit{Fr}$ is the Froude number, which quantifies 
the ratio of inertial to gravitational (i.e., buoyancy) forces. From this order-of-magnitude estimate, it follows that when $\mathit{Fr} \ll 1$, $W\ll U$, i.e., the motion is tangential. This indicates that there are strong restoring buoyancy forces opposing radial motion. This condition for tangential motions of gravity waves can be fulfilled in the regions of clusters we are considering \citep[as shown in][who measured $\mathit{Fr}<0.2$ for $r<0.2\rtc$ from a hydrodynamical cluster simulation]{valdarnini2019}.

Although the preceding discussion applies strictly to short-wavelength modes within the WKBJ approximation, numerical simulations of the nonlinear evolution of global g-modes in galaxy clusters demonstrate that these modes can still be effectively understood using insights from linear WKBJ analysis. \cite{lufkin1995} showed this for an individual galaxy as a source of gravity waves, while \cite{kim2007} conducted an analysis for multiple galaxies. 

How can we connect these considerations to the magnetic fields? The evolution of magnetic fields in the ideal MHD case is governed by the induction equation:
\begin{equation}
\frac{\partial \vec{B}}{\partial t} = \nabla \times (\vec{v} \times \vec{B})\, .
\end{equation}
If we assume idealized spherical symmetry in the limit of dominant tangential motion, that is, $v_r \ll v_\theta, v_\phi$, the induction equation in $(r,\theta,\phi)$ simplifies to:
\begin{equation}
\frac{\partial \vec{B}}{\partial t} = \left(0,  \ \frac{1}{r} \frac{\partial}{\partial r} (r v_\theta B_r), \ \frac{1}{r} \frac{\partial}{\partial r} (r v_\phi B_r)\right)^{\mathrm{T}}\, .
\end{equation}
This implies that there is no or only weak evolution in the radial magnetic field component, while there is evolution in the tangential magnetic field components. The evolution of these tangential magnetic fields is driven by tangential flows $v_\theta, v_\phi$. The stronger the radial dependence, the stronger the effect.

These theoretical considerations highlight that in a stably stratified medium, gravity waves can be excited and trapped within specific regions. Gravity waves are preferentially tangential
causing the magnetic fields to also be preferentially tangential within those regions.
As noted in Section~\ref{velocity_anisotropy}, while tangential motions are necessary to establish tangential magnetic fields, a mere lack of tangential velocity bias at a given point in time does not imply that magnetic fields cannot be tangential. Magnetic fields can indeed remain tangential even after the velocity field has decayed or randomized (randomization of velocity directions does necessarily imply randomization of the magnetic fields that retain long-term memory of plasma motions).

As noted above, trapping of gravity modes is associated with the generation of vorticity. While the equation governing the evolution of vorticity is mathematically similar to the ideal MHD induction equation for the magnetic fields, these equations are not identical because of the presence of the baroclinic term in the vorticity evolution equation ($\propto \vec{\nabla}P \times \vec{\nabla}\rho$). Consequently, the evolution of vorticity and magnetic fields does not have to be exactly the same. In fact, generation of vorticity that tracks the g-modes (see Eq.~\ref{eq:vorticity_generation}) is specifically attributed to the presence of the baroclinic term in the vorticity evolution equation \citep[see Eq. 5 in][]{lufkin1995}.

\subsubsection{Trapping of gravity waves in TNG-Cluster}

We now assess the possibility of gravity wave trapping in our cosmological simulations. Fig.~\ref{fig:wBVvsBetaA} shows the median profiles of the Brunt-Väisälä frequency $\omega_{BV}$, as defined in Eq.~\ref{eq:wBV}, for SCCs (blue), WCCs (purple), and NCCs (orange; all solid lines) in TNG-Cluster at $z=0$. For comparison of the relevant radial regimes, we also show the median profiles of the magnetic anisotropy (dashed lines; y axis on the right). The horizontal gray dotted lines indicate three arbitrary values for the stirring frequency $\omega_{\rm stir} = v_{\rm turb}/l$. These values are chosen to showcase the consequences of different perturbing/stirring frequencies. 

We find all three $\omega_{BV}$ median profiles are highest in the core and lowest at $\rvir$. At those extrema, clusters have similar values, regardless of CC status. However, at radii $\sim 0.03-0.2$, the shape of the profiles differs noticeably. $\omega_\text{BV}$ in SCCs has a distinctive bump at these radii, while the profile for NCCs is more flat. The profile of WCCs is intermediate between the two, with a shape similar to the median NCC profile.

This analysis shows that SCCs generally have higher $\omega_\text{BV}$ than NCCs at $0.1r_{500}$. This makes trapping of gravity waves easier in SCCs. In particular, we find that trapping is possible for SCCs in the range of $\omega_{\rm stir} = 5-10 \,\rm{Gyr}^{-1}$. Lower stirring frequencies would enable trapping in WCCs and NCCs as well, when considering the population as a whole. In addition, it is suggestive that the minimum in $\betaA$ of SCCs is aligned with the bump in $\omega_{BV}$, albeit somewhat shifted to smaller radii. That is, the region where we find tangentially oriented magnetic fields is similar to the radial range where the trapping of gravity waves can occur. In this scenario AGN feedback, mergers and satellite halos stir the ICM and thereby excite the gravity waves. Regardless of the origin of perturbations, this suggests that the ICM conditions specific to cool-core clusters -- namely, their degree of entropy stratification -- lead to an efficient reshaping of magnetic fields into preferentially tangential orientations.

\subsubsection{Redshift evolution}

We have seen that the population-wide dip in $\beta_A$ for SCCs vanishes at higher redshifts (Fig.~\ref{fig:betaA_mass_z}). In particular, the population-wide dip in $\beta_A$ disappears by $z\gtrsim0.5$. To be consistent with the gravity wave trapping picture, we expect that the bump in the $\omega_{BV}$ profiles for SCCs should diminish towards earlier times, causing the behavior of SCCs, WCCs, and NCCs to converge. 

To investigate this, we consider the evolution of the $\omega_\text{BV}$ profiles at $z>0$ (not shown). We find that the median $\omega_{BV}$ profile for NCCs remains largely unchanged across redshifts, while the median profile for WCCs gradually approaches that of NCCs. For SCCs, the bump in the median $\omega_{BV}$ profile decreases with increasing redshift, where this decrease begins at larger radii. By $z=0.4$, there is only a minimal population-wide dip in $\beta_A$, coinciding with a small remaining bump in $\omega_{BV}$ above $\omega_{\text{stir}} \sim 10 \, \text{Gyr}^{-1}$ at the same radius. Notably, the central increase (towards $r=0$) in $\omega_{BV}$ persists at all redshifts for both SCCs and WCCs. The turning point, where values begin to rise, shifts to slightly larger radii as redshift increases.\footnote{When examining redshift trends for SCCs based on following their direct progenitors from $z=0$, we find similar results. The dip in $\beta_A$ shifts to slightly larger radii with increasing redshift and disappears by $z=0.7$. Similarly, the bump in $\omega_{BV}$ decreases over time, with its maximum moving to larger radii as redshift increases.}

This consistent redshift evolution supports the idea that tangential magnetic fields are linked to trapped gravity waves, with a stirring frequency of order $\sim 5-10 \, \text{Gyr}^{-1}$.

\subsubsection{Situation in the core}

\begin{figure*}
    \centering
    \includegraphics[width=0.33\textwidth]{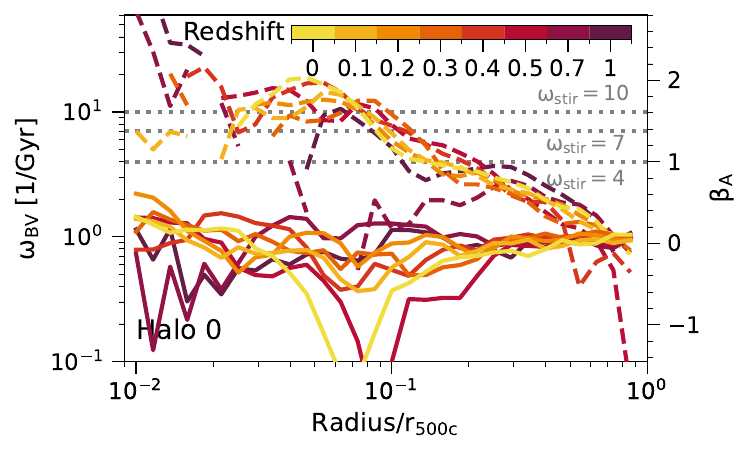}
    \includegraphics[width=0.33\textwidth]{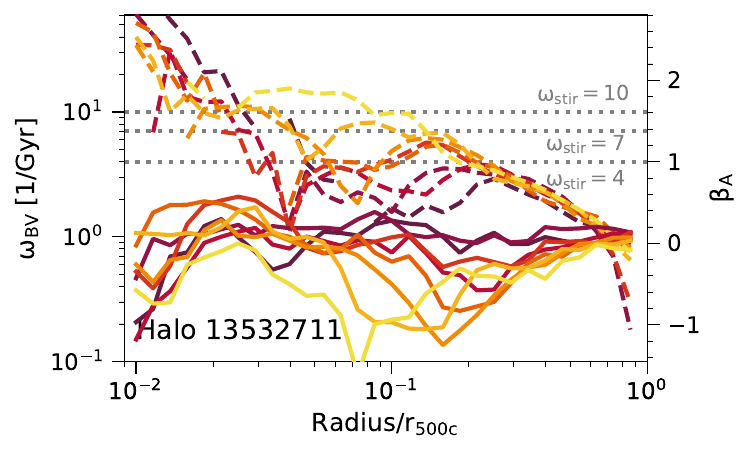}
    \includegraphics[width=0.33\textwidth]{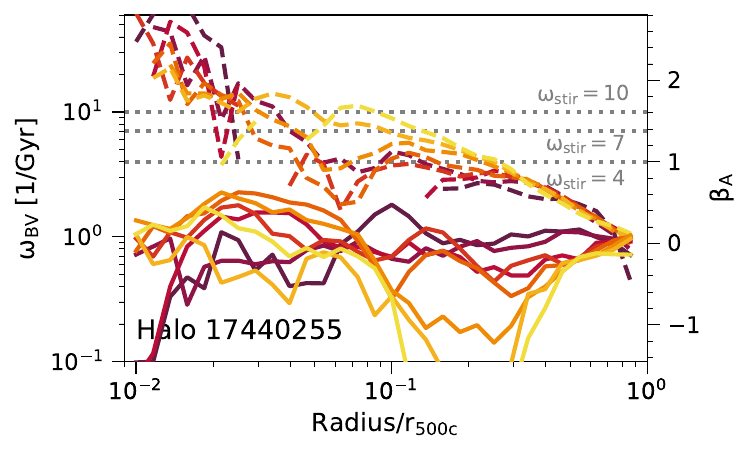}
    \includegraphics[width=0.33\textwidth]{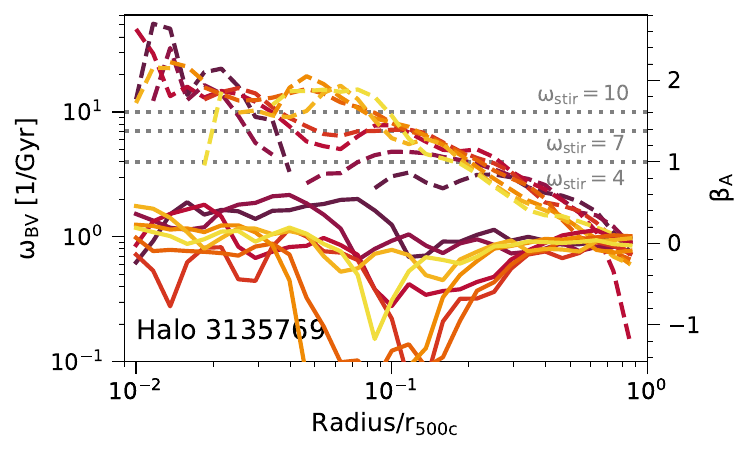}
    \includegraphics[width=0.33\textwidth]{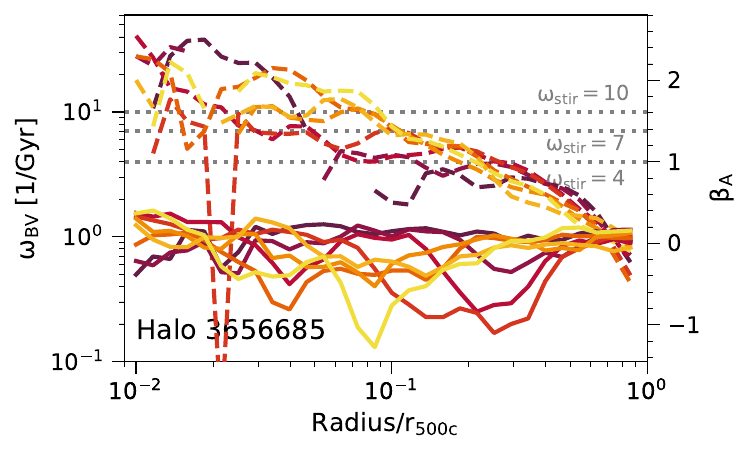}
    \includegraphics[width=0.33\textwidth]{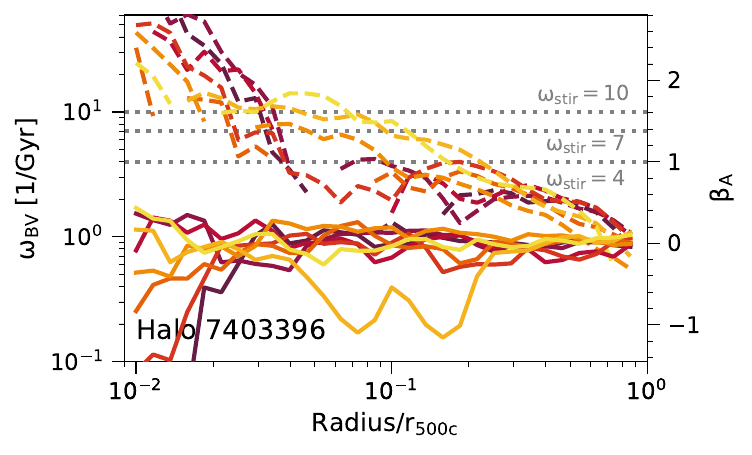}
    \caption{Redshift evolution of the magnetic anisotropy $\betaA$ and Brunt-Väisälä profiles for six example halos with pronounced dips in $\betaA$ from TNG-Cluster. There is considerable diversity in their evolution: dips can exist at $z=0$, vanish and re-appear at higher redshifts, grow less weak and eventually vanish, or first grow stronger before later vanishing. There are also clusters that do not have a dip in $\betaA$ at $z=0$, but harxad one at higher redshift, while the radial position of the dips can itself move with redshift.}\label{fig:timeEvo}
\end{figure*}

A puzzle remains: given that the Brunt-Väisälä frequency is largest in the core, implying that gravity waves should be most easily trapped, why is there no corresponding tangential feature in cluster core magnetic fields?

First, the stirring frequency may be higher in the core, such that the condition $\omega_{\rm stir} < \omega_{BV}$ is not fulfilled. That is, although g-wave trapping could occur, a sufficiently short time-scale stirring process will prevent such behavior. In particular, AGN feedback is the likely culprit. To assess this, we make a rough measurement of the duty-cycle, i.e., the time between discrete energy injection events for kinetic mode SMBH feedback in the TNG model, characteristic of high-mass cluster halos at low redshift. We find $\omega_{BV}\sim 20\, \mathrm{Gyr}^{-1}$ and $\omega_{\rm AGN}\sim \omega_{\rm stir}\sim 13\, \mathrm{Gyr}^{-1}$ in the core.\footnote{This value is based on \cite{lehle2025}, where we measure the time difference between individual AGN events in a single prototypical TNG-Cluster halo, where high time resolution outputs are available. Using a set of 100 TNG-Cluster BCGs at $z=0$ which reside in SCCs and WCCs and taking the interval between the last 3 energy injection for each, \cite{prunier2025a} find a even higher value of $ \omega_{\rm stir}\sim 28\, \mathrm{Gyr}^{-1}$.} Both frequencies are comparable, indicating that gravity wave trapping is likely inhibited in the core. In some sense, AGN feedback can destroy any coherent structure that would otherwise form due to the trapping of gravity waves.

Another possible reason that we do not find tangential magnetic fields in the core, despite the higher values of \(\omega_\text{BV}\), lies in the limitations of linear theory. The derivation of the gravity wave dispersion relation in Eq.~\ref{eq:wBV_disprel} relies on the WKBJ approximation. However, this approximation may break down in the core, where radii are small. Referring back to Eq.~\ref{eq:W/U=Fr}, if the characteristic length scale \(L\) decreases as \(r \rightarrow 0\), the ratio \(W/U\) would increase. By definition, gravity waves are only tangential when \(W \ll U\), such that the conditions in the core are less favorable to gravity wave trapping.

Other astrophysical causes may also play a role. For example, sloshing motions can induce non-negligible relative motions between dark matter and gas in the center of clusters. These speeds can reach appreciable fractions of the sound speed, meaning that it is difficult for the gas to rapidly respond. The orientation of magnetic fields with respect to the current location of the cluster center, as defined by the gravitational potential minimum, will then not necessarily appear tangential. As a result, gentle sloshing motions could erase any tangential bias close to CC cores, while leaving such structure largely unaffected at larger radii. 

\subsection{Timescales of formation and existence of tangentially biased magnetic fields.}

How fast can the tangentially biased magnetic fields form, and how long do they persist? We make a preliminary assessment of these timescales by considering the evolution of individual halos. Fig.~\ref{fig:timeEvo} shows the redshift evolution of the radial profiles of magnetic anisotropy and BV frequency for six SCC clusters. We select these halos to showcase a broad diversity in the evolution of tangentially biased magnetic fields.

First, several halos have tangential magnetic fields that persist over long time periods: the dip in $\beta$ can exist stably for $\sim 5\,$Gyr up to $z=0.4$ (upper middle panel). It tends to move to larger radii at higher redshift, and this is also true for the bump in $\omega_{BV}$. On the other hand, there are examples where the location of the tangential magnetic field feature does not move (upper right panel). In addition, there are many halos where we identify a dip at only one snapshot. For instance, existing only at $z=0$ but vanishing at earlier times but reappearing at even higher redshift (upper left panel). In this case, these are the two redshifts where $\omega_{BV}$ has the highest value, although the differences of $\omega_{BV}$ at different times are relatively small. Another example shows a dip only at $z=0.1$, but this is not the redshift of the highest bump in $\omega_{BV}$ (lower right panel). Finally, we see an example of a cluster where a dip feature exists at $z=0$, disappears, and then reappears at $z=0.2-0.5$ (lower left panel).

Overall, we find that tangential magnetic field features can persist for either short or long time periods, from at least 1 Gyr to 5 Gyr. Their formation can be as fast as $\sim 150$ Myr, which is our nomimal time spacing between individual snapshots in TNG-Cluster. This is of order $1 / \omega_{BV} \sim 0.1$\,Gyr. The diversity of timescales we identify may reflect multiple origins, i.e., stirring mechanisms, or simply that ICM conditions for gravity wave trapping can be stable, or rapidly evolving, depending on the conditions in a particular cluster.

In a cosmological simulation there are unavoidably many sources of perturbations, all present, that could initially excite gravity waves. Most notably, both AGN feedback as well as interactions and mergers with other halos. All of our TNG-Cluster halos experience multiple instances of both types of perturbations at late times. Moreover, we have shown that the resulting preferential magnetic field orientation is in agreement with the theory of trapped gravity waves. However, we identify this phenomenon only in cool-core clusters. Such systems also have relatively high AGN activity, and while the two are potentially casually linked, they may also reflect the same underlying cause, such as recent accretion and merger activity that enhances the gas density in the cluster core. 

\section{Summary and Conclusions} \label{sec_conclusions}

In this paper we study the properties and morphology of magnetic fields using 352 galaxy clusters from the new TNG-Cluster cosmological magnetohydrodynamical simulation. We focus on the orientation of magnetic fields in CC versus NCC halos. 
Our main findings are:

\begin{itemize}
    \item \textbf{The intracluster medium is weakly magnetized.} We find field strengths of $1-30\mu\,$G in the central ICM of TNG-Cluster halos. The magnetic field strength increases with halo mass and toward the core. The highest central magnetic field strengths are found in SCC clusters. (Figs.~\ref{fig:galleryB} and \ref{fig:profilesB})
    \item \textbf{Magnetic fields in cluster progenitors are rapidly amplified at $z>2$.} Peak field strengths are reached at cosmic noon, and can subsequently decline or remain roughly constant depending on distance and evolutionary history. (Fig.~\ref{fig:BvsM500vsZ})
    \item \textbf{Magnetic fields are dynamically subdominant.}
    Typical values of $\betaP\sim 200-2000$ imply that the hot, volume-filling gas is dominated by thermal pressure. SCC clusters have notably smaller $\betaP$ at the center as well as larger $\betaP$ at $r\sim0.1\rvir$, while WCCs and NCCs have similar values across radii. (Figs.~\ref{fig:profileBetaP} and \ref{fig:galleryBetaP}) 
    \item \textbf{Magnetic fields in SCCs are tangentially biased.} We find negative values of magnetic anisotropy preferentially among SCCs. This feature is localized at a characteristic distance of $\sim 0.1 \rvir$. Among NCCs and WCCs, there is no such population-wide feature, where magnetic fields instead tend towards isotropy. (Fig.~\ref{fig:betaA}) 
    \item \textbf{Mass and redshift trends of anisotropy.} There is no halo mass dependence for the tangential magnetic field configurations. In addition, the unique tangential feature we identify in SCCs occurs only at low redshifts $z \lesssim 0.4$ and vanishes at higher redshift. (Fig.~\ref{fig:betaA_mass_z})
    \item \textbf{The tangential bias of magnetic fields is consistent with gravity wave trapping.} We discuss several scenarios that may explain the origin of the reported field configurations. We find evidence that in individual halos either AGN feedback or merger shocks and sloshing motions could source initial perturbations. However, given that preferentially tangential magnetic fields appear only in cool-core clusters, we ultimately conclude that it is their unique entropy stratification (gradients) that enable the trapping of gravity waves and thus the reorientation of magnetic fields into preferentially tangential topologies at $\sim 0.1 \rvir$. (Fig.~\ref{fig:wBVvsBetaA})
    \item \textbf{Tangential features can form quickly, but also persist over several Gyrs.} Different clusters show a diversity of evolutionary pathways. In some cases, the feature we identify is long-lived, while in others it is transitory or even quasi-reoccurring. This reflects a diversity of stirring mechanisms and/or a diversity in the entropy structure of cool-core clusters. (Fig.~\ref{fig:timeEvo})
\end{itemize}

Thanks to TNG-Cluster, in this paper, we have identified a rather unique feature -- that cool-core clusters have a tendency toward preferential magnetic field alignment in the tangential, i.e., azimuthal directions at characteristic scales of $\sim 0.1 \rvir$. We show that this phenomenon is consistent with a simple theoretical picture related to gravity wave trapping. In this scenario AGN feedback, mergers and satellite halos can all act as a source of perturbation that excite gravity waves. At the same time, recent observations of clusters may find tentative evidence of similar field configurations. In the future, it will be useful to make quantitative predictions for the observable signatures of such magnetic field topologies. Additional simulations and analysis that carefully control for the effects of AGN feedback, merger history, and ICM entropy structure will be essential to pin down the ultimate origin of this phenomenon.

\section*{Data Availability}

The IllustrisTNG simulations, including TNG-Cluster, are publicly available and accessible at \url{www.tng-project.org/data} \citep{nelson2019a}. Data directly related to this publication is available on request from the corresponding authors.

\begin{acknowledgements}
KL acknowledges funding from the Hector Fellow Academy through a Research Career Development Award. MR acknowledges support from the National Aeronautics and Space Administration grant ATP 80NSSC23K0014 and the National Science Foundation Collaborative Research grant NSF AST-2009227. DN acknowledges funding from the Deutsche Forschungsgemeinschaft (DFG) through an Emmy Noether Research Group (grant number NE 2441/1-1). KL is a Fellow of the International Max Planck Research School for Astronomy and Cosmic Physics at the University of Heidelberg (IMPRS-HD). AP acknowledges funding from the European Union (ERC, COSMIC-KEY, 101087822, PI: Pillepich). MP acknowledges funding from the Physics Department of the University of Montreal (UdeM) and the Centre for Research in Astrophysics of Quebec (CRAQ). KL would like to thank Wonki Lee for insightful discussions that helped to shape our discussion. The TNG-Cluster simulation has been carried out on several machines: as part of the TNG-Cluster project on the HoreKa supercomputer, funded by the Ministry of Science, Research and the Arts Baden-Württemberg and by the Federal Ministry of Education and Research. The bwForCluster Helix supercomputer, supported by the state of Baden-Württemberg through bwHPC and the German Research Foundation (DFG) through grant INST 35/1597-1~FUGG. The Vera, Cobra, and Raven clusters of the Max Planck Computational Data Facility (MPCDF). The BinAC cluster, supported by the High Performance and Cloud Computing Group at the Zentrum für Datenverarbeitung of the University of Tübingen, the state of Baden-Württemberg through bwHPC and the German Research Foundation (DFG) through grant no INST 37/935-1~FUGG. This analysis has been carried out on the Vera supercomputer of the Max Planck Institute for Astronomy (MPIA).
\end{acknowledgements}

\bibliographystyle{aa}
\bibliography{refs}

\end{document}